\documentclass[aps,physrev,twocolumn,superscriptaddress]{revtex4}
\usepackage{graphicx}
\usepackage{dcolumn}
\usepackage{bm}
\usepackage{epsfig}
\usepackage{amsmath}
\input{epsf}
\usepackage{psfrag}
\usepackage{comment}
\usepackage{xcolor}
\usepackage{float}
\usepackage{braket}
\usepackage{appendix}
\usepackage{booktabs}
\usepackage{multirow}
\usepackage{subcaption}
\usepackage{slashed}
\usepackage{bbm}
\allowdisplaybreaks[4]
\newcommand{\bea}{\begin{eqnarray}}
\newcommand{\eea}{\end{eqnarray}}

\begin{document}
\title{Helicity Amplitudes and Quantum Entanglement Analysis in Fully Charm Tetraquark Decays into Double Vector Charmonia or Charmed Mesons}
\author{Kaiwen Chen$^{1}$, Feng-Xiao Liu$^{2,3}$\footnote{liufx@ihep.ac.cn}, Qiang Zhao$^{3,4}$\footnote{zhaoq@ihep.ac.cn}, Xian-Hui Zhong$^{5}$\footnote{zhongxh@hunnu.edu.cn}, Ruilin Zhu$^{1,6,7}$\footnote{rlzhu@njnu.edu.cn}, Bing-Song Zou$^{6,7,8,9}$\footnote{zoubs@itp.ac.cn}}
\affiliation{
$^1$ Department of Physics and Institute of Theoretical Physics, Nanjing Normal University, Nanjing, Jiangsu 210023, China\\
$^2$School of Nuclear Science and Technology, University of South China, Hengyang, 421001, Hunan,
China\\
$^3$ Institute of High Energy Physics, Chinese Academy of Sciences, Beijing 100049, China\\
$^4$  University of Chinese Academy of Sciences, Beijing 100049, China \\
$^5$ Department of Physics, Hunan Normal University, and Key Laboratory of Low-Dimensional
Quantum Structures and Quantum Control of Ministry of Education, Changsha 410081, China\\
$^6$ CAS Key Laboratory of Theoretical Physics, Institute of Theoretical Physics,
Chinese Academy of Sciences, Beijing 100190, China\\
$^7$ Peng Huanwu Innovation Research Center, Institute of Theoretical Physics,
Chinese Academy of Sciences, Beijing 100190, China\\
$^8$ Department of Physics, Tsinghua University, Beijing 100084, China\\
$^9$ Southern Center for Nuclear-Science Theory (SCNT), Institute of Modern Physics,
Chinese Academy of Sciences, Huizhou 516000, China}

\begin{abstract}

We study helicity amplitudes for the fully charm tetraquark states decays into vector meson pair from single gluon exchange contribution.   The decay angular distributions have been given by the cascade decays $T_{4c}\to J/\psi(D_{(s)}^*)+J/\psi(\bar{D}_{(s)}^*)$ along with $J/\psi\to \mu^++\mu^-$ or $D_{(s)}^*\to D_{(s)}+\pi$, showing that spin-0 and spin-2 states can be distinguished.  We also find that the spin-0 state decay to vector meson pair exhibits a higher degree of quantum entanglement than that in spin-2 state decays. These findings will help distinguish among different spin-parity states in experiment and explore possible as-yet-undiscovered hadronic structures, thereby shedding light on the internal properties of the double-$J/\psi$ exotic states discovered at the LHC. 
\end{abstract}

\maketitle

\section{Introduction}

Fifty-two years ago, the discovery of the $J/\psi$ meson by teams led by Samuel Ting at Brookhaven National Laboratory~\cite{E598:1974sol} and Burton Richter at SLAC~\cite{SLAC-SP-017:1974ind} provided direct evidence for the existence of the fourth quark, i.e. the charm quark, and opened the door to heavy-flavour physics. This groundbreaking achievement, historically referred to as the ``November Revolution" was a pivotal moment in the development of the Standard Model of particle physics. Now we have entered the era of exploring fully heavy hadorn states composed of at least two charm quarks and two anti-charm quarks.

In 2020, a narrow structure around 6.9 GeV in the invariant mass spectrum of double $J/\psi$ was discovered using the proton-proton collision data at centre-of-mass energies of $\sqrt{s}=$7, 8 and 13 TeV recorded by the LHCb experiment at the Large Hadron Collider, corresponding to an integrated luminosity of $9~fb^{-1}$~\cite{LHCb:2020bwg}. Subsequently, this exotic structure was confirmed by other two independent experiments ATLAS~\cite{ATLAS:2023bft} and CMS~\cite{CMS:2023owd}, and new exotic structures were also discovered in these experiments. Both ATLAS and CMS experimental data indicated the possibility of fully charm tetraquark family in the mass region from 6.2 to 7.3 GeV. In the double $J/\psi$ spectrum, three exotic states with masses around 6.6, 6.9, and 7.1 GeV are reported~\cite{CMS:2023owd,CMS:2026tiu}. Furthermore,  the 6.9 GeV structure has also been discoverd in the \(J/\psi\)-\(\psi(2S)\) invariant mass distribution by the ATLAS Collaboration~\cite{ATLAS:2025nsd}.
In a recent study, the spin-parity numbers for these states are more likely to be $J^{PC}=2^{++}$ assuming that all these exotic states have the same spin-parity assignment  by the CMS Collaboration~\cite{CMS:2025fpt}. 

Theoretical studies of fully heavy tetraquark states prior to the experimental discovery can be found in these references~\cite{Iwasaki:1975pv,Chao:1980dv,Anwar:2017toa,Karliner:2016zzc,Debastiani:2017msn,Wu:2016vtq,Liu:2019zuc,Wang:2019rdo,Bedolla:2019zwg,Wang:2017jtz,Wang:2018poa,Heupel:2012ua,Esposito:2018cwh}.
The major theoretical approaches on fully charm tetraquark mass spectrum include quark potential models~\cite{Jin:2020jfc,Liu:2021rtn,Faustov:2020qfm,Lu:2020cns,Wang:2021kfv,Wu:2025wvv},  QCD sum rules~\cite{Chen:2016jxd,Zhang:2020xtb,Wang:2020dlo,Wang:2020ols,Albuquerque:2020hio,Wang:2021mma,Wan:2020fsk,Yang:2020wkh,Agaev:2023wua,Agaev:2023gaq,Tang:2024zvf,Tang:2024kmh,Tang:2025ept},  Bethe-Salpeter equations~\cite{Li:2021ygk,Ke:2021iyh}, lattice QCD calculations~\cite{Li:2025vbd,Li:2025ftn,Meng:2024czd},  the diquark-antidiquark cluster model~\cite{Zhu:2020xni,Giron:2020wpx,lodha:2025snn}, the tetraquark-molecule mixing schemes~\cite{Santowsky:2021bhy},  and dynamically generated mechanism~\cite{Wang:2020wrp,Gong:2020bmg,Dong:2020nwy,Guo:2020pvt,Huang:2024jin}. The production cross section of fully charm tetraquarks can be found in Ref.~\cite{Zhu:2020xni,Liu:2025mxv,Feng:2023agq,Belov:2024qyi,Celiberto:2025vra,Zhang:2020hoh,Feng:2020riv}. The complete next-to-leading order QCD calculation with resummation effects has been completed in~\cite{Wang:2025hex}. 

The decay process of fully charm tetraquark $T_{4c}\to J/\psi(D_{(s)}^*)+J/\psi(\bar{D}_{(s)}^*)$ along with $J/\psi\to \mu^++\mu^-$ or $D_{(s)}^*\to D_{(s)}+\pi$ are of particular interest, as such cascade decays provide access to the angular variables of the final-state particles, which encode the spin information of the initial tetraquark, the polarization of the intermediate vector mesons and the possible exotic structures below double-$J/\psi$ threshold. Pioneering frameworks for analyzing such multi-dimensional angular distributions via helicity amplitudes can be found in Ref.~\cite{Jacob:1959at}.

In this work,  we focus only on double vector charmonia  and double vector charmed mesons transitions of the S-wave fully charm tetraquark states with spin 0 and 2 which can directly couple to the exotic states appearing in the double $J/\psi$ spectrum, while physical states with other quantum numbers from orbitally excitation will be considered in the future. The content of the paper is as follows: In section II, we derive in detail the analytical expressions for the angular distributions of the cascade decays $T_{4c} \to J/\psi+ J/\psi$ and $T_{4c} \to D_{(s)}^* +\bar{D}_{(s)}^*$ along with $J/\psi\to \mu^++\mu^-$ or $D_{(s)}^*\to D_{(s)}+\pi$. In section III, we introduce the Quark Model (QM) and Heavy Quark Spin Symmetry (HQSS) to calculate the the helicity amplitudes, and then give the numerical results for the angular distributions for the above cascade decays. In the section IV, we analyzed the entanglement of the two final-state vector mesons through the decay of a fully charm tetraquark. A brief conclusion will be given in the end.

\section{Helicity Decomposition and Angular Distribution Formulae}
In this section,  we derive carefully the four body decay angular distributions of fully charm tetraquark by the cascade decays $T_{4c}\to J/\psi+J/\psi$ along with two separate charmonium leptonic decay $J/\psi\to \mu^++\mu^-$, which have also measured 
from the latest CMS experiments.  The cascade decay angular distributions of $T_{4c}\to D_{(s)}^*+\bar{D}_{(s)}^*$ along with $D_{(s)}^*\to D_{(s)}+\pi$ can be otained similarly.

The S-matrix element for the cascade decay $T_{4c} \to J/\psi J/\psi \to 2\mu^+2\mu^-$ can be constructed by inserting the completeness relation for the intermediate resonance,
\begin{align}
	\mathbbm{1}&=\sum_{s} \int \frac{d^4p}{(2\pi)^4}\frac{i| J/\psi(p,s) \rangle \langle J/\psi(p,s) |}{p^2-m_{J/\psi}^2-im_{J/\psi}\Gamma_{J/\psi}}. 
\end{align}
For the decay process $T_{4c} \to J/\psi J/\psi \to 2\mu^+2\mu^-$, using $H_1(x)$, $H_2(y)$ and $H_3(z)$ to denote the effective Hamiltonians of decays $T_{4c} \to J/\psi J/\psi$ and two kinds of $J/\psi \to \mu^+\mu^-$ respectively, the S-matrix can be written as
\begin{align}
	&\langle  \mu^+\mu^-\mu^+\mu^- | S | T_{4c} \rangle=\langle  \mu^+\mu^-\mu^+\mu^-| iT | T_{4c} \rangle  \nonumber \\
	=&(-i)^3\int d^4x d^4y d^4z \langle \mu^+\mu^-\mu^+\mu^- |\mathcal{H}_1(x) \mathcal{H}_2(y)\mathcal{H}_3(z)| T_{4c} \rangle \nonumber \\
	=&-i\sum_{\lambda_1,\lambda_2} \int d^4x d^4y d^4z \frac{d^4p_{1}}{(2\pi)^4}\frac{d^4p_{2}}{(2\pi)^4}\langle J/\psi J/\psi|\mathcal{H}_1(x)| T_{4c} \rangle \nonumber \\
	&\times \frac{i\langle \mu^+\mu^-|\mathcal{H}_2(y)| J/\psi \rangle}{p_{1}^2-m^2_{J/\psi}-im_{J/\psi}\Gamma_{J/\psi}}\frac{i\langle \mu^+\mu^-|\mathcal{H}_3(z)| J/\psi \rangle}{p_{2}^2-m^2_{J/\psi}-im_{J/\psi}\Gamma_{J/\psi}} \nonumber \\
	=&i\sum_{\lambda_1,\lambda_2}(2\pi)^4\delta^4(p_{H}-p_{\mu1}-p_{\mu2}-p_{\mu3}-p_{\mu4}) \nonumber \\
	&\times \mathcal{A}(T_{4c} \to J/\psi J/\psi) \prod_{i=1}^{2} \frac{\mathcal{A}_i(J/\psi \to \mu^+\mu^-)}{p_i^2-m_{J/\psi}^2-im_{J/\psi}\Gamma_{J/\psi}}.
\end{align}
So the total decay amplitude can  be expressed as
\begin{align}
	&i\mathcal{A}(T_{4c} \to J/\psi J/\psi \to 4\mu)=i\sum_{\lambda_1,\lambda_2}\mathcal{A}(T_{4c} \to J/\psi J/\psi)  \nonumber \\
	&\times \frac{\mathcal{A}_1(J/\psi \to \mu^+\mu^-)}{p_1^2-m_{J/\psi}^2-im_{J/\psi}\Gamma_{J/\psi}}\frac{\mathcal{A}_2(J/\psi \to \mu^+\mu^-)}{p_2^2-m_{J/\psi}^2-im_{J/\psi}\Gamma_{J/\psi}},
\end{align}
where $\lambda_1$ and $\lambda_2$ are the helicities of the two intermediate $J/\psi$ mesons. 

According to the Jacob-Wick angular formalism~\cite{Jacob:1959at}, the helicity amplitude for a generic two-body decay process $a \to b+c$ can be decomposed with the Wigner D-matrix element as~\cite{Jacob:1959at}
\begin{align}
	\langle \Omega, \lambda_b, \lambda_c | \mathcal{H} | J_a, \lambda_a \rangle = \sqrt{\frac{2J_a + 1}{4\pi}} D^{J_a *}_{\lambda_a, \lambda_b - \lambda_c}(\Omega) F^{J_a}_{\lambda_b \lambda_c},
\end{align}
where $J_a$ and $\lambda_a$ are the spin and its projection of particle $a$, $\lambda_b$ and $\lambda_c$ are the helicities of the $b$ and $c$ particles, $\Omega$ describes the solid angle of the decay products in the rest frame of $a$, and $F^{J_a}_{\lambda_b \lambda_c}$ is the reduced helicity amplitude containing the dynamic information of the interaction. 
In our scenario, assuming the initial $T_{4c}$ state is unpolarized, we must average over its initial spin projections $\lambda_a$. Then, we integrate over the production solid angle $\Omega_0$ of the $T_{4c} \to J/\psi J/\psi$ process. Utilizing the unitarity and orthogonality of the Wigner D-matrices, we have
\begin{align}
	\sum_m \int d\Omega_0 D^{J*}_{m,\Delta\lambda}(\Omega_0) D^J_{m,\Delta\lambda'}(\Omega_0)  =\frac{8\pi^2}{2J+1} \delta_{\Delta\lambda, \Delta\lambda'}.
\end{align} 
 Consequently, the intermediate spin density matrix elements $h_{\lambda'_1 \lambda'_2}^{\lambda_1 \lambda_2}$ for the decay $T_{4c} \to J/\psi J/\psi$ can be defined purely in terms of the helicity amplitudes:
\begin{equation}
	h_{\lambda'_1 \lambda'_2}^{\lambda_1 \lambda_2} = F^{J_a}_{\lambda_1 \lambda_2} F_{\lambda'_1 \lambda'_2}^{J_a*}.
\end{equation}
Combining the above formulas, we obtain the squared modulus of the four-body decay amplitude as
\begin{align}
	|\mathcal{A}|^2=&\sum_{\lambda_1,\lambda_2,\lambda_1^\prime,\lambda_{a_i},\lambda_{b_i}}\frac{N_{s_{J/\psi}}^2D^{*s_{J/\psi}}_{\lambda_1^\prime,\lambda_{a}}D^{s_{J/\psi}}_{\lambda_1,\lambda_a}|F^\prime_{\lambda_{a_1}\lambda_{a_2}}|^2}{(p_1^2-m_{J/\psi}^2)^2+m_{J/\psi}^2\Gamma_{J/\psi}^2} \nonumber \\
	& \times \frac{N_{s_{J/\psi}}^2D^{*s_{J/\psi}}_{\lambda_2^\prime,\lambda_{b}}D^{s_{J/\psi}}_{\lambda_2,\lambda_b}|F^{\prime\prime}_{\lambda_{b_1}\lambda_{b_2}}|^2}{(p_2^2-m_{J/\psi}^2)^2+m_{J/\psi}^2\Gamma_{J/\psi}^2}h^{\lambda_1\lambda_2}_{\lambda_1^\prime\lambda_2^\prime}.
\end{align}
In the above equation, $N_{s_{J/\psi}}=\sqrt{\frac{2s_{J/\psi}+1}{4\pi}}$, $\lambda'_2=\lambda'_1-\lambda_{1}+\lambda_{2}$, $\lambda_a=\lambda_{a_1}-\lambda_{a_2}$, $\lambda_b=\lambda_{b_1}-\lambda_{b_2}$. And $\lambda_{a_1}, \lambda_{a_2}, \lambda_{b_1}, \lambda_{b_2}$ denotes the helicities of the four $\mu$. Meanwhile, we denote the helicity amplitudes for the two $J/\psi \to \mu^+\mu^-$ decays by $F^\prime_{\lambda_{a_1}\lambda_{a_2}}$ and $F^{\prime\prime}_{\lambda_{b_1}\lambda_{b_2}}$. For the entire process, its partial decay width can be expressed as
\begin{align}
	d\Gamma=&\sum_{\lambda_1,\lambda_2,\lambda_1^\prime,\lambda_{a_i},\lambda_{b_i}}\frac{(2\pi)^4}{2M_T}\frac{1}{2s_T+1}h^{\lambda_1\lambda_2}_{\lambda_1^\prime\lambda_2^\prime}\nonumber \\
	&\times\frac{N_{s_{J/\psi}}^2D^{*s_{J/\psi}}_{\lambda_1^\prime,\lambda_{a}}D^{s_{J/\psi}}_{\lambda_1,\lambda_a}|F_{\lambda_{a_1}\lambda_{a_2}}|^2}{(p_1^2-m_{J/\psi}^2)^2+m_{J/\psi}^2\Gamma_{J/\psi}^2}\nonumber\\
	& \times \frac{N_{s_{J/\psi}}^2D^{*s_{J/\psi}}_{\lambda_2^\prime,\lambda_{b}}D^{s_{J/\psi}}_{\lambda_2,\lambda_b}|F_{\lambda_{b_1}\lambda_{b_2}}|^2}{(p_2^2-m_{J/\psi}^2)^2+m_{J/\psi}^2\Gamma_{J/\psi}^2}d\Phi_4.
\end{align}
For the four-body phase space, we can write it as
\begin{align}
	d\Phi_4&=\int\delta^4(p_H-p_{\mu1}-p_{\mu2}-p_{\mu3}-p_{\mu4})\sum_{i=1}^4 \frac{d^3p_{\mu i}}{(2\pi)^32E_i} \nonumber  \\
	&= d\Phi_2(T_{4c} \to J/\psi J/\psi)d\Phi_2(J/\psi \to \mu^+\mu^-)\nonumber\\
	&~~~\times d\Phi_2(J/\psi \to \mu^+\mu^-)dm_{12}^2dm_{34}^2  \nonumber \\
	&=\frac{1}{(2\pi)^{12}}\int\frac{|p_H|d\Omega_0}{4M_T}\frac{|p_{\mu1}|d\Omega_1}{4m_{12}}\frac{|p_{\mu3}|d\Omega_2}{4m_{34}}dm_{12}^2dm_{34}^2. 
\end{align}

The total decay width for the fully charm tetraquark decay into double $J/\psi$ can be written as
\begin{align}
	\Gamma(T_{4c} \to J/\psi J/\psi)=&\frac{(2\pi)^4}{2M_T}\frac{\sum_{\lambda_1\lambda_2}|F_{\lambda_1\lambda_2}|^2}{2s_T+1} \nonumber \\
	&\times \int d\Phi_2(T_{4c} \to J/\psi J/\psi)\nonumber\\
	=&\frac{|p_H|}{8\pi M_T^2}\frac{\sum_{\lambda_1\lambda_2}|F_{\lambda_1\lambda_2}|^2}{2s_T+1}
	.
\end{align}

\begin{figure}[th]
	\includegraphics[width=0.5\textwidth]{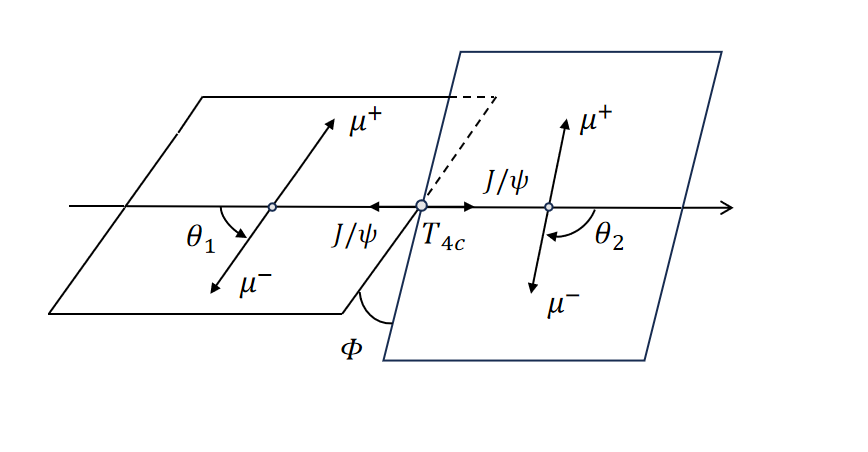}
	\caption{The illustration  of helicity angles in fully charm tetraquark decays into four muons. $\theta_i$ with $i=1,2$ is the angle between one muon momentum in the rest frame of $J/\psi$ relative to the $J/\psi$ momentum in the rest frame of fully charm tetraquark. $\Phi$ is the angle between two secondary decay planes.}\label{Figangles}
\end{figure}

For the remaining part, we define a normalized angular distribution function $ W(\theta_1, \theta_2, \Phi)$ as in Ref.~\cite{Kramer:1991xw}.  Thus, the fully differential decay width $d\Gamma$ can be factorized into the two-body production width and a normalized joint angular distribution $W(\theta_1, \theta_2, \Phi)$
\begin{align}
	&d\Gamma = \Gamma(T_{4c} \to J/\psi J/\psi)\times W(\theta_1, \theta_2, \Phi) d\Omega_1 d\Omega_2,
\end{align}
where $\theta_i$ and $\Phi$ are the polar and azimuthal angles as defined in Fig.~\ref{Figangles}. Without loss of generality, we set the azimuthal angle of one of the $J/\psi \to \mu^+\mu^-$ decay planes to zero, making $\Phi$ the relative angle between the two decay planes.

The normalized angular distribution $W(\theta_1, \theta_2, \Phi)$ is obtained by contracting the production density matrix with the secondary decay density matrices $A$ and $B$:
\begin{align}
	W(\theta_1, \theta_2, \Phi) = \mathcal{N} \sum_{\lambda_1, \lambda_2} ^{\lambda'_1, \lambda'_2} h^{\lambda_1 \lambda_2}_{\lambda'_1 \lambda'_2} A_{\lambda_1 \lambda'_1}(\theta_1, \Phi) B_{\lambda_2 \lambda'_2}(\theta_2, 0), \label{w}
\end{align}
where $\mathcal{N}$ is the global normalization factor ensuring $\int W d\Omega_1 d\Omega_2 = {\cal B}^2$ in the four body cascade decays with the leptonic branching ratio ${\cal B}={\cal B}(J/\psi \to \mu^+\mu^-)$. Taking $A_{\lambda_1 \lambda'_1}$ as an example, it is defined as:
\begin{align}
	A_{\lambda_1 \lambda'_1}(\theta_1, \Phi) = \frac{\sum_{\lambda_{a_1}, \lambda_{a_2}} R_{\lambda_1 \lambda'_1}(\theta_1, \Phi, \lambda_{a_1}, \lambda_{a_2})}{\sum_{\lambda_{a_1}, \lambda_{a_2}} |T(p_1, \lambda_{a_1}, \lambda_{a_2})|^2},
\end{align}
where $T(p_1, \lambda_{a_1}, \lambda_{a_2})$ is the reduced dynamic matrix element for the $J/\psi$ decay. The term $R_{\lambda_1 \lambda'_1}$ encapsulates the geometric Wigner D-matrix products for the secondary decay, which is defined as:
\begin{align}
	&R_{\lambda_1 \lambda'_1}(\theta_1, \Phi, \lambda_{a1}, \lambda_{a2}) = \frac{3}{4\pi} |T(p_1, \lambda_{a_1}, \lambda_{a_2})|^2 \nonumber \\
	&\times D^{1*}_{\lambda_1, \lambda_{a_1} - \lambda_{a_2}}(\Phi, \theta_1, -\Phi) \times D^{1}_{\lambda'_1, \lambda_{a_1} - \lambda_{a_2}}(\Phi, \theta_1, -\Phi). 
\end{align}
The definition of $B_{\lambda_i \lambda'_j}(\theta, 0)$ is exactly the same as $A_{\lambda_i \lambda'_j}(\theta, \Phi)$ but evaluated at $\Phi = 0$. By substituting these matrices into Eq.~\ref{w} and summing over the helicity states, the complete angular distribution of the cascade decay is obtained .

In our paper, we only consider the S-wave tetaquarks with spin-parity $J^{PC}=0^{++}, 2^{++}$ here, which can directly couple to the exotic states appearing in the double $J/\psi$ spectrum. After the above calculation, the general decay angular distribution of $T_{4c}\left(0^{++}\right)  \rightarrow J/\psi(\mu^+\mu^-)+ J/\psi(\mu^+\mu^-)$ is

\begin{widetext}
\begin{align}
	\frac{d^3\Gamma}{d\cos\theta_1 d\cos\theta_2 d\Phi}=\frac{9p_{J/\psi}{\cal B}^2}{256\pi^2 M_T^2}
	&\Big\{ \sin^2\theta_1\sin^2\theta_2h^{00}_{00}+\frac{1}{4}
	(1+\cos^2\theta_1)(1+\cos^2\theta_2)(h^{11}_{11}+h^{-1-1}_{-1-1}) \nonumber \\
	&+\frac{1}{8}\sin^2\theta_1\sin^2\theta_2[\cos2\Phi Re(h^{11}_{-1-1})-\sin2\Phi Im(h^{11}_{-1-1})] \nonumber \\
	&+\frac{1}{8}\sin2\theta_1\sin2\theta_2[\cos\Phi Re(h^{11}_{00}+h^{-1-1}_{00})-\sin\Phi Im(h^{11}_{00}-h^{-1-1}_{00})] \Big\}.
\end{align}
\end{widetext}
And for the case of $T_{4c}\left(2^{++}\right)  \rightarrow J/\psi(\mu^+\mu^-)+ J/\psi(\mu^+\mu^-)$, the angular distribution can be expressed as
\begin{widetext}
\begin{align}
	\frac{d^3\Gamma}{d\cos\theta_1d\cos\theta_2d\Phi}=
	&\frac{9p_{J/\psi}{\cal B}^2}{5120\pi^2M_T^2}\Big\{ 2\sin^2\theta_1\sin^2\theta_2 h^{00}_{00}+\frac{1}{2}(1+\cos^2\theta_1)(1+\cos^2\theta_2)(h^{-1-1}_{-1-1}+h^{11}_{11}+h^{1-1}_{1-1}\nonumber \\
	&+h^{-11}_{-11})+(1+\cos^2\theta_1)\sin^2\theta_2(h^{10}_{10}+h^{-10}_{-10})+\sin^2\theta_1(1+\cos^2\theta_2)(h^{01}_{01}+h^{0-1}_{0-1})\nonumber \\
	&+\frac{1}{2}\sin2\theta_1\sin2\theta_2[\cos\Phi Re(h^{11}_{00}+h^{00}_{-1-1}-h^{10}_{0-1}-h^{01}_{-10})-\sin\Phi Im(h^{11}_{00}+h^{00}_{-1-1}-h^{10}_{0-1}-h^{01}_{-10})]\nonumber \\
	&+\sin^2\theta_1\sin^2\theta_2[\cos2\Phi Re(h^{11}_{-1-1})-\sin2\Phi Im(h^{11}_{-1-1})\Big\}.
\end{align}
\end{widetext}
Note that the above angular distribution  only depend on the helicity amplitude in the first step decay of fully charm tetraquark into two charmonium, since the lepton mass is ignored and then the dynamic information in the secondary step decays becomes trivial.

In addition to fully charm tetraquark decay into double $J/\psi$, we also calculate the angular distributions of fully charm tetraquark into double charmed mesons, as they collectively determine the tetraquark's inner structure. In the case of $T_{4c}\left(0^{++}\right)  \rightarrow D^*(\rightarrow D \pi)+ \bar{D}^*(\rightarrow \bar{D} \pi)$, the general decay angular distribution is
\begin{widetext}
	\begin{align}
		\frac{d^3\Gamma}{d\cos\theta_1 d\cos\theta_2 d\Phi}=\frac{9p_{D^*}{\cal B^\prime}^2}{64 \pi^2 M_T^2} & \Big \{ \cos^2\theta_1\cos^2\theta_2h^{00}_{00}+\frac{1}{4}\sin^2\theta_1\sin^2\theta_2(h^{11}_{11}+h^{-1-1}_{-1-1}) \nonumber \\
		&+\frac{1}{2}\sin^2\theta_1\sin^2\theta_2[\cos2\Phi Re(h^{11}_{-1-1})-\sin2\Phi Im(h^{11}_{-1-1})] \nonumber \\
		&+\frac{1}{4}\sin2\theta_1\sin2\theta_2 [\cos\Phi Re(h^{11}_{00}+h^{-1-1}_{00})-\sin\Phi Im(h^{11}_{00}-h^{-1-1}_{00})] \Big \}.
	\end{align}
\end{widetext}
Therein $\cal B^\prime$ is branching ratio for $D^* \to D\pi$. And for the case of $T_{4c}\left(2^{++}\right)  \rightarrow D^*(\rightarrow D \pi)+ \bar{D}^*(\rightarrow \bar{D} \pi)$, the angular distribution becomes
\begin{widetext}
\begin{align}
	\frac{d\Gamma}{d\cos\theta_1d\cos\theta_2d\Phi}=
	&\frac{9p_{D^*}{\cal B^\prime}^2}{1280\pi^2 M_T^2}\Big\{2\cos^2\theta_1\cos^2\theta_2h^{00}_{00}+\frac{1}{2}\sin^2\theta_1\sin^2\theta_2(h^{11}_{11}+h^{-1-1}_{-1-1}+h^{1-1}_{1-1}+h^{-11}_{-11}) \nonumber\\
	&+\sin^2\theta_1\cos^2\theta_2(h^{10}_{10}+h^{-10}_{-10})+\cos^2\theta_1\sin^2\theta_2(h^{01}_{01}+h^{0-1}_{0-1}) \nonumber\\
	&+\frac{1}{2}\sin2\theta_1\sin2\theta_2[\cos\Phi Re(h^{01}_{01}+h^{0-1}_{0-1}+h^{11}_{00}+h^{00}_{-1-1} -h^{10}_{0-1}-h^{01}_{-10}) \nonumber\\
	&-\sin\Phi Im(h^{01}_{01}+h^{0-1}_{0-1}+h^{11}_{00}+h^{00}_{-1-1}-h^{10}_{0-1}-h^{01}_{-10})] \nonumber\\
	&+\sin^2\theta_1\sin^2\theta_2[\cos2\Phi Re(h^{11}_{-1-1})-\sin2\Phi Im(h^{11}_{-1-1})]\Big\}.
\end{align}
\end{widetext}
Note that the above angular distributions are also only related to the helicity amplitude in the first step decay of fully charm tetraquark into charmed meson pair. The reason is because the pion spin is zero and then the dynamic information in the secondary step decays also becomes trivial.
For the case of $T_{4c}\left(0^{++}\right)  \rightarrow D_s^*(\rightarrow D_s \pi)+ \bar{D}_s^*(\rightarrow \bar{D}_s \pi)$, the general decay angular distribution can be also obtained by the replacement of $D^*\to D^*_s$  and $D\to D_s$.

It should be also noted that in many physical scenarios, the decay process exhibits additional symmetries that impose strict relations among these amplitudes. Consequently, the general angular distribution can be greatly simplified under such symmetry constraints, leading to more compact and predictive expressions. 

For the decay  process with two identical final particles, the helicity amplitude needs to satisfy the following symmetry, 
\begin{align}
F^{J}_{\lambda_1\lambda_2}=(-1)^JF^{J}_{\lambda_2\lambda_1}.\label{relation1}
\end{align}
Also, if parity is conserved in decay process, there is another relation between helicity amplitudes
\begin{align}
F^{J}_{\lambda_1\lambda_2}=\eta\eta_1\eta_2(-1)^{J-s_1-s_2}F^{J}_{-\lambda_1-\lambda_2},\label{relation2} 
\end{align}
where $\eta$ and $\eta_i$ are the parity of the involving particles.

Since the two decay processes satisfy the conservation of parity or the final state consists of two identical particles, the formula Eqs.~\ref{relation1} and \ref{relation2} can be used to simplify the angular distribution. 

Using these constraints, the simplified decay angular distributions for $T_{4c}\left(0^{++}\right) \rightarrow J/\psi(\rightarrow \mu^+\mu^-) + J/\psi(\rightarrow \mu^+\mu^-)$ and $T_{4c}\left(2^{++}\right) \rightarrow J/\psi(\rightarrow \mu^+\mu^-) + J/\psi(\rightarrow \mu^+\mu^-)$ are,
\begin{widetext}
 \begin{align}\label{angudis1}
\frac{d^3\Gamma(0^{++})}{d\cos\theta_1 d\cos\theta_2 d\Phi}=&\frac{9p_{J/\psi}{\cal B}^2}{256\pi^2 M_T^2}
\Big\{ h^{00}_{00}\sin^2\theta_1\sin^2\theta_2+\frac{1}{2}h^{11}_{11}
(1+\cos^2\theta_1)(1+\cos^2\theta_2)+\frac{1}{8}h^{11}_{11}\sin^2\theta_1\sin^2\theta_2\cos2\Phi \nonumber \\
&+\frac{1}{8} h^{11}_{00}\sin2\theta_1\sin2\theta_2\cos\Phi \Big\},
 \end{align}
and
	\begin{align}\label{angudis2}
		\frac{d^3\Gamma(2^{++})}{d\cos\theta_1d\cos\theta_2d\Phi}=&\frac{9p_{J/\psi}{\cal B}^2}{10240\pi^2M_T^2} \Big\{2h_{00}^{00}+9h_{11}^{11}+12h_{-10}^{-10}+9h_{-11}^{-11}+(-2h_{00}^{00}+3h_{11}^{11}-4h_{-10}^{-10}+3h_{-11}^{-11})\cos2\theta_2  \nonumber \\
		&+(-2h_{00}^{00}+3h_{11}^{11}-4h_{-10}^{-10}+3h_{-11}^{-11})\cos2\theta_1+(2h_{00}^{00}+h_{11}^{11}-4h_{-10}^{-10}+h_{-11}^{-11})\cos2\theta_1\cos2\theta_2 \nonumber \\
		& +4h_{11}^{11}\cos2\Phi\sin^2\theta_1\sin^2\theta_2+4(h_{00}^{11}-h_{-10}^{-10})\cos\Phi\sin2\theta_1\sin2\theta_2)  \Big\},
	 \end{align}
	 respectively.

The simplified decay angular distributions for $T_{4c}\left(0^{++}\right) \rightarrow D^*(\rightarrow D\pi) + \bar{D}^*(\rightarrow \bar{D}\pi)$ and $T_{4c}\left(2^{++}\right) \rightarrow D^*(\rightarrow D\pi) + \bar{D}^*(\rightarrow \bar{D}\pi)$ are,

	\begin{align}
		\frac{d^3\Gamma(0^{++})}{d\cos\theta_1 d\cos\theta_2 d\Phi}=&\frac{9p_{D^*}{\cal B^\prime}^2}{64 \pi^2 M_T^2}  \Big \{ \cos^2\theta_1\cos^2\theta_2h^{00}_{00}+\frac{1}{2}\sin^2\theta_1\sin^2\theta_2h^{11}_{11}  +\frac{1}{2}\sin^2\theta_1\sin^2\theta_2\cos2\Phi h^{11}_{11} \nonumber \\
		&+\frac{1}{2}\sin2\theta_1\sin2\theta_2\cos\Phi h^{11}_{00} \Big \},
	\end{align}
	and,	
	\begin{align}
		\frac{d\Gamma(2^{++})}{d\cos\theta_1d\cos\theta_2d\Phi}=
		&\frac{9p_{D^*}{\cal B^\prime}^2}{1280\pi^2M_T^2}\Big\{2\cos^2\theta_1\cos^2\theta_2h^{00}_{00}+\sin^2\theta_1\sin^2\theta_2(h^{11}_{11}+h^{1-1}_{1-1})+2\sin^2\theta_1\cos^2\theta_2h^{10}_{10} \nonumber\\
		&+2\cos^2\theta_1\sin^2\theta_2h^{01}_{01}+\sin2\theta_1\sin2\theta_2\cos\Phi (h^{11}_{00}+h^{01}_{01}-h^{10}_{01}) +2\sin^2\theta_1\sin^2\theta_2\cos2\Phi h^{11}_{11}\Big\},
	\end{align}
	respectively.
\end{widetext}

\section{Helicity Amplitudes Caculation}

The next core step is to calculate the helicity amplitudes for the two-body decays of fully charm tetraquarks. Once we have this dynamical information, we can immediately obtain the four-body decay angular distributions in above.

In order to comprehensively display the dominant decay modes of fully charm tetraquark, we list the two-body channels in the following:
\begin{align}\label{decaychannel}
T_{4c}\to H_{c\bar{c}}+H^{(\prime)}_{c\bar{c}},&\;\;\;T_{4c}\to H_{c\bar{q}}+H'_{q\bar{c}},\nonumber\\
T_{4c}\to g+g,\;\;\;\;\;\;\;\;\;&\;\;\; T_{4c}\to T'_{4c}+\gamma \nonumber \\
T_{4c}\to T'_{4c}+h,\;\;\;\;\;\;&\;\;\;T_{4c}\to H_{c\bar{c}}+h,\nonumber\\
T_{4c}\to H_{c\bar{c}}+\gamma,\;\;\;\;\;&\;\;\;T_{4c}\to \gamma+\gamma. 
\end{align}

 \begin{figure}[th]
        \includegraphics[width=0.45\textwidth]{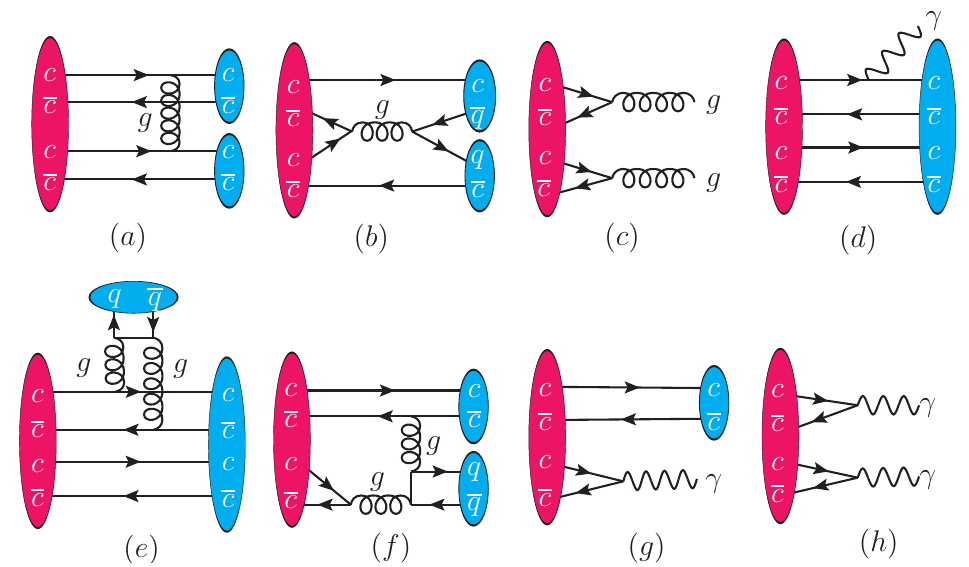}
        \caption{Typical Feynman diagrams for fully charm tetraquark two-body decays, illustrating six kinds of modes: double charmonia transition (a), double charmed meson transition (b), two-gluon annihilation (c), electromagnetic transitions (d, g), light meson transitions (e, f), and two-photon annihilation (h). }\label{Figmechanism}
  \end{figure}
These decay channels can be classified into six kinds: double charmonia transition,
double charmed meson transition,  two-ihilatigluon annon, electromagnetic transition, light meson transition,  and two-photon annihilation. The typical Feynman diagrams are shown in Fig.~\ref{Figmechanism}. Therein the contributions of  double charmonia transition and double charmed meson transition are expected to be domimant to determine the total decay width and other diagram contributions are suppressed according to the magnitude of the coupling strength. And the two-gluon annihilation decay width has been predicted to be less than $0.1MeV$ in Ref.~\cite{Sang:2023ncm}  and become smaller 
after inputting the latest constraint of the 
nonperturbative matrix elments and including the QCD corrections in Ref.~\cite{Wang:2026asl}. In this paper, we only consider the helicity amplitude calculation for double charmonia transition through quark model and double charmed meson transition for fully charm tetraquarks through heavy quark spin symmetry. The helicity analysis for other decay channels will be put in future work.

In practice, the commonly used ones are the 
helicity amplitudes and the partial-wave amplitudes. For two-body decay, they are respectively defined as
\begin{align}
	&F^J_{\lambda_1\lambda_2}=4\pi \sqrt{\frac{m}{p}}\langle J\lambda\lambda_1\lambda_2|\mathcal{H}| J\lambda \rangle,  \nonumber \\
	&a^J_{ls}=4\pi \sqrt{\frac{m}{p}}\langle J\lambda ls|\mathcal{H}| J\lambda\rangle, 
\end{align}
where 
$m$ is the mass of the parent particle, and 
$p$ is the magnitude of the momentum of the final-state particle. Since the helicity amplitudes and the partial-wave amplitudes form two complete orthogonal bases, we can also express the partial-wave amplitudes in terms of the helicity amplitudes,
\begin{align}
	a^J_{ls}=\sum_{\lambda_{1}\lambda_{2}}\sqrt{\frac{2l+1}{2J+1}}F^J_{\lambda_1\lambda_2}\langle l0s\lambda|J\lambda\rangle \langle s_1\lambda_1s_2-\lambda_2|s\lambda\rangle.
\end{align}
Conversely, the helicity amplitudes $F^J_{\lambda_1\lambda_2}$ can be expressed by the formula
\begin{align}
	F^J_{\lambda_1\lambda_2}=\sum_{ls}\sqrt{\frac{2l+1}{2J+1}}a^J_{ls}\langle l0s\lambda|J\lambda\rangle \langle s_1\lambda_1s_2-\lambda_2|s\lambda\rangle.
\end{align}
In the above formula, $l$ represents the orbital angular momentum quantum number, $s$ represents the total spin of the two daughter particles, $s_1$ and $s_2$ denote the spins of the two daughter particles, $\lambda_1$ and $\lambda_2$ are their respective helicities, and $J$ and $\lambda$ are the spin of the parent particle and its helicity.

On the other hand, the decay amplitude is usually expressed in terms of independent Lorentz invariants.  For a particle with spin zero into two vector mesons, its decay amplitude can be expressed as \cite{Kramer:1991xw}
\begin{align}
	&\mathcal{A}(0^{++}\to 1^-+1^-)\nonumber\\
	=&\epsilon_{1 \mu}^* \epsilon_{2 \nu}^*\left(a g^{\mu \nu}+\frac{b p^\nu_1 p^\mu_2}{m_1 m_2} +\frac{i c \epsilon^{\mu \nu \alpha \beta} p_{1 \alpha} p_\beta}{m_1 m_2} \right), \label{3}
\end{align}
where $m_i$, $p_i$ and $\epsilon_i$ are the mass, momentum and polarization vector for the two daughter particles, respectively. The relation between the parameters a,b,c and the helicity amplitudes can be expressed as
\begin{align}
	&F^0_{11}=a + \sqrt{x^2-1}c , \nonumber \\
	&F^0_{-1-1}=a -\sqrt{x^2-1}c,\nonumber \\
	&F^0_{00}=-ax-b(x^2-1) ,
\end{align}
where
\begin{align}
	x^2=\frac{p_m^2M_T^2}{m_1^2m_2^2}+1.
\end{align}
In the above formula, $p_m$ represents the momentum of the decaying particle in the rest frame of the parent particle and $M_T$ represents the mass of the parent particle. The coefficient $c$ vanishes when the decay is parity conserved. This is consistent with the helicity relations given by Eqs.~\ref{relation1} and \ref{relation2}.

For a particle with spin two into two vector mesons, its decay amplitude can be written as \cite{Gao:2010qx}
\begin{align}
&\mathcal{A}(2^{++}\to 1^-+1^-)
\nonumber\\
=&\epsilon_1^{*\mu}\epsilon_2^{*\nu}[c_1p_1\cdot p_2\epsilon_{3\mu\nu}+c_2 g_{\mu \nu} \epsilon_{3\alpha \beta} \tilde{p}^\alpha \tilde{p}^\beta\nonumber \\&+c_3 \frac{p_{2 \mu} p_{1 \nu}}{M_T^2} \epsilon_{3\alpha \beta} \tilde{p}^\alpha \tilde{p}^\beta +c_5 \epsilon_{3\alpha \beta} \frac{\tilde{p}^\alpha \tilde{p}^\beta}{M_T^2} \epsilon_{\mu \nu \rho \sigma} p_1^\rho p_2^\sigma\nonumber \\
	&+2 c_4(p_{1 \nu} p_2^\alpha \epsilon_{3\mu \alpha}+p_{2 \mu} p_1^\alpha \epsilon_{3\nu \alpha}) \nonumber \\
	&+\frac{c_7 \epsilon_3^{\alpha \beta} \tilde{p}_\beta}{M^2}\left(\epsilon_{\alpha \mu \rho \sigma} p^\rho \tilde{p}^\sigma p_\nu+\epsilon_{\alpha \nu \rho \sigma} p^\rho \tilde{p}^\sigma p_\mu\right)\nonumber \\
	&+c_6 \epsilon_3^{\alpha \beta} \tilde{p}_\beta \epsilon_{\mu \nu \alpha \rho} p^\rho], \label{4}
\end{align}
where $\epsilon_3$ represents the polarization tensor of a tensor particle and $\tilde{p}=p_1-p_2$. The relation between helicity amplitude and coefficient $c_i$ can be obtained by the following formula
\begin{align}
	& F^2_{1-1}=F^2_{-11}=\frac{M_T^2}{4} c_1\left(1+\beta^2\right), \nonumber \\
	& F^2_{11}=\frac{M_T^2}{\sqrt{6}}\left[\frac{c_1}{4}\left(1+\beta^2\right)+2 c_2 \beta^2+i \beta\left(c_5 \beta^2-2 c_6\right)\right], \nonumber \\
	& F^2_{-1-1}=\frac{M_T^2}{\sqrt{6}}\left[\frac{c_1}{4}\left(1+\beta^2\right)+2 c_2 \beta^2-i \beta\left(c_5 \beta^2-2 c_6\right)\right], \nonumber \\
	& F^2_{10}=F^2_{01}=\frac{M_T^3}{m \sqrt{2} }\left[\frac{c_1}{8}\left(1+\beta^2\right)+\frac{c_4}{2} \beta^2-\frac{c_6+c_7 \beta^2}{2} i \beta\right], \nonumber \\
	& F^2_{-10}=F^2_{0-1}=\frac{M_T^3}{m \sqrt{2} }\left[\frac{c_1}{8}\left(1+\beta^2\right)+\frac{c_4}{2} \beta^2+\frac{c_6+c_7 \beta^2}{2} i \beta\right], \nonumber \\
	& F^2_{00}=\frac{M_T^4}{m^2 \sqrt{6} }\left[\left(1+\beta^2\right)\left(\frac{c_1}{8}-\frac{c_2}{2} \beta^2\right)-\beta^2\left(\frac{c_3}{2} \beta^2-c_4\right)\right],
\end{align}
where $m$ stands for the mass of the daughter particle and $\beta=\sqrt{1-\frac{4m^2}{M_T^2}}$. For the two processes, we find  that the coefficients $c_5$, $c_6$, and $c_7$ vanish if parity is conserved. This is also consistent with the relation given by 
Eqs.~\ref{relation1} and \ref{relation2}.

In the following, we will calculate the decay helicity amplitudes of fully charm tetraquark states $T_{4c}$ with spin-zero and spin-two into two vector mesons, where
two methods, i.e.,  quark model and heavy quark spin symmetry, are employed.

\subsection{Quark model}

We first introduce how to calculate the helicity amplitude for $T_{4c} \to J/\psi J/\psi$ with quark model. In quark model, the four-quark system $cc\bar{c}\bar{c}$ obeys the Pauli principle and color confinement and has  no other restrictions. The masses and wave functions of the initial fully charm tetraquark
states  are taken from the constituent quark model
calculation of Ref.~\cite{liu:2020eha}. The Hamiltonian adopted in Ref.~\cite{liu:2020eha} contains the
full kinetic-energy and potential-energy terms, with the kinetic-energy
term treated in the nonrelativistic form. The potential part for any two charm quarks in four-body system is expanded as
\begin{equation}
V_{ij}=V_{ij}^{conf}+V_{ij}^{coul}+V_{ij}^{SS}+V_{ij}^{LS}+V_{ij}^{T}.
\end{equation}
The confinement potential $V_{ij}^{conf}$, Coulomb potential $V_{ij}^{Coul}$,
spin-spin contact hyperfine potential $V_{ij}^{SS}$, spin-orbit
potential $V_{ij}^{LS}$, and the tensor term $V_{ij}^{T}$ are standard in quark potential model as~\cite{Godfrey:1985xj}
\begin{eqnarray}
	V_{ij}^{conf} & = & -\frac{3}{16}\left(\boldsymbol{\lambda}_{i}\cdot\boldsymbol{\lambda}_{j}\right)\left(b_{ij}r_{ij}\right),\label{vcon}
\end{eqnarray}
\begin{eqnarray}
	V_{ij}^{Coul} & = & \frac{1}{4}\left(\boldsymbol{\lambda}_{i}\cdot\boldsymbol{\lambda}_{j}\right)\left(\frac{\alpha_{ij}}{r_{ij}}\right),\label{vcon-1}
\end{eqnarray}
\begin{eqnarray}
	V_{ij}^{SS} & = & -\frac{\alpha_{ij}}{4}\left(\boldsymbol{\lambda}_{i}\cdot\boldsymbol{\lambda}_{j}\right)\left\{ \frac{\sigma_{ij}^{3}e^{-\sigma_{ij}^{2}r_{ij}^{2}}}{\pi^{3/2}}\frac{8(\mathbf{S}_{i}\cdot\mathbf{S}_{j})}{3m_{i}m_{j}}\right\},~~~~\label{voge ss}
\end{eqnarray}
\begin{eqnarray}
	V_{ij}^{LS} & = & -\frac{\alpha_{ij}}{16}\frac{\left(\boldsymbol{\lambda}_{i}\cdot\boldsymbol{\lambda}_{j}\right)}{r_{ij}^{3}}\bigg(\frac{1}{m_{i}^{2}}+\frac{1}{m_{j}^{2}}+\frac{4}{m_{i}m_{j}}\bigg)
	LS_+\nonumber \\
	&  & -\frac{\alpha_{ij}}{16}\frac{\left(\boldsymbol{\lambda}_{i}\cdot\boldsymbol{\lambda}_{j}\right)}{r_{ij}^{3}}\bigg(\frac{1}{m_{i}^{2}}-\frac{1}{m_{j}^{2}}\bigg)LS_-,
\end{eqnarray}
\begin{eqnarray}
	V_{ij}^{T} & = & -\frac{\alpha_{ij}}{4}\left(\boldsymbol{\lambda}_{i}\cdot\boldsymbol{\lambda}_{j}\right)\cdot\frac{1}{m_{i}m_{j}r_{ij}^{3}}
	\nonumber\\&&\times\Bigg\{\frac{3(\mathbf{S}_{i}\cdot\mathbf{r}_{ij})(\mathbf{S}_{j}\cdot\mathbf{r}_{ij})}{r_{ij}^{2}}-\mathbf{S}_{i}\cdot\mathbf{S}_{j}\Bigg\}.\label{voge ten}
\end{eqnarray}
In the above equations, $r_{ij}\equiv|\mathbf{r}_{i}-\mathbf{r}_{j}|$
is the distance between the $i$-th and $j$-th quarks, and $\boldsymbol{\lambda}_{i,j}$ are the color operators acting
on the $i,j$-th quarks. $LS_+=\mathbf{L}_{ij}\cdot(\mathbf{S}_{i}+\mathbf{S}_{j})$ and $LS_-=\mathbf{L}_{ij}\cdot(\mathbf{S}_{i}-\mathbf{S}_{j})$.
$\mathbf{L}_{ij}$ represents
the relative orbital angular momentum between the $i$-th and $j$-th
quarks, and $\mathbf{S}_{i}$
represents the spin of the $i$-th quark. The parameters $b_{ij}$ and $\alpha_{ij}$ denote the strength
of the confinement and the strong coupling of the one-gluon-exchange
potential, respectively. All these  parameters in quark model can be found in Ref.~\cite{liu:2020eha}. Specifically, there are four types of quark interactions $V_{12}$,
$V_{34}$, $V_{14}$, and $V_{23}$ that drive the decay of the tetraquark
state. 

The decay amplitude $\mathcal{M}(A\to BC)$ is then given by
\begin{equation}
\begin{aligned}
\mathcal{M}(A\to BC) = &-\sqrt{(2\pi)^3}
\sqrt{8M_A E_B E_C}\\
&\times \left\langle BC\left|
V_{12}+V_{34}+V_{14}+V_{23}
\right|A\right\rangle,
\end{aligned}
\end{equation}
where the interactions $V_{13}$ and $V_{24}$ within 
the two final-state mesons are already included in 
their respective Hamiltonians and are therefore 
excluded from the transition operator. The remaining 
four interactions, $V_{12}$, $V_{34}$, $V_{14}$, 
and $V_{23}$, drive the quark rearrangement decay.
Here, $A$ represents the initial tetraquark state, 
and $BC$ represents the final hadron pair.
$M_{A}$ is the mass of the initial state. The mass of the initial state is the physical mass eigenvalue obtained after diagonalizing the mass matrix. $E_{B}$ and $E_{C}$ are the energies of the final states $B$ and $C$, respectively. The energies of the final-state hadrons can be completely determined by the two-body kinematic relations as
\begin{align}
	&E_B=\sqrt{M_B^2+p_f^2}, \quad E_C=\sqrt{M_C^2+p_f^2}, \nonumber \\
	&p_f=\frac{\sqrt{[M_A^2-(M_B+M_C)^2][M_A^2-(M_B-M_C)^2]}}{2M_A}.
\end{align}
For simplicity, the wave functions of the $A$, $B$, and $C$ hadron states are parametrized in the form of a single harmonic oscillator.

\begin{table}[htp]
	\caption{\label{ConventionI} Configurations for the tetraquark $cc\bar{c}\bar{c}$ system considered  from quark model in this work. The last column lists the masses of the physical tetraquark states obtained after configuration mixing~\cite{liu:2020eha}.}
	\begin{tabular}{lcccc}
		\hline
		\hline
		Configuration  & \multicolumn{2}{c}{ Wave Function } &  Mass(MeV)& \tabularnewline
		\hline
		$1^{1}S_{0^{++}(6\bar{6})_{c}}$  & \multicolumn{1}{c}{$\psi_{000}^{1S}\chi_{00}^{00}$} & $|6\bar{6}\rangle^{c}$ & 6455&  \tabularnewline
		$1^{1}S_{0^{++}(\bar{3}3)_{c}}$  & \multicolumn{1}{c}{$\psi_{000}^{1S}\chi_{00}^{11}$} & $|\bar{3}3\rangle^{c}$ &  6550& \tabularnewline
		$1^{5}S_{2^{++}(\bar{3}3)_{c}}$  & \multicolumn{1}{c}{$\psi_{000}^{1S}\chi_{22}^{11}$} & $|\bar{3}3\rangle^{c}$ &6524 & \tabularnewline
		\hline
		$2^{1}S_{0^{++}(6\bar{6})_{c}\left(\xi_{1},\xi_{2}\right)}$  & \multicolumn{1}{c}{$\sqrt{\frac{1}{2}}\left(\psi_{100}^{\xi_{1}}+\psi_{100}^{\xi_{2}}\right)\chi_{00}^{00}$} & $|6\bar{6}\rangle^{c}$ & 6908& \tabularnewline
		$2^{1}S_{0^{++}(\bar{3}3)_{c}\left(\xi_{1},\xi_{2}\right)}$  & \multicolumn{1}{c}{$\sqrt{\frac{1}{2}}\left(\psi_{100}^{\xi_{1}}+\psi_{100}^{\xi_{2}}\right)\chi_{00}^{11}$} & $|\bar{3}3\rangle^{c}$ & 6957& \tabularnewline
		$2^{1}S_{0^{++}(6\bar{6})_{c}\left(\xi_{3}\right)}$  & \multicolumn{1}{c}{$\psi_{100}^{\xi_{3}}\chi_{00}^{00}$} & $|6\bar{6}\rangle^{c}$ & 7018& \tabularnewline
		$2^{1}S_{0^{++}(\bar{3}3)_{c}\left(\xi_{3}\right)}$  & \multicolumn{1}{c}{$\psi_{100}^{\xi_{3}}\chi_{00}^{11}$} & $|\bar{3}3\rangle^{c}$ & 7185& \tabularnewline
		$2^{5}S_{2^{++}(\bar{3}3)_{c}\left(\xi_{1},\xi_{2}\right)}$  & \multicolumn{1}{c}{$\sqrt{\frac{1}{2}}\left(\psi_{100}^{\xi_{1}}+\psi_{100}^{\xi_{2}}\right)\chi_{22}^{11}$} & $|\bar{3}3\rangle^{c}$ &6927 & \tabularnewline
		$2^{5}S_{2^{++}(\bar{3}3)_{c}\left(\xi_{3}\right)}$  & \multicolumn{1}{c}{$\psi_{100}^{\xi_{3}}\chi_{22}^{11}$} & $|\bar{3}3\rangle^{c}$ & 7032& \tabularnewline
		\hline
		\hline
	\end{tabular}
\end{table}

\begin{figure}[htp]
	\includegraphics[width=0.45\textwidth]{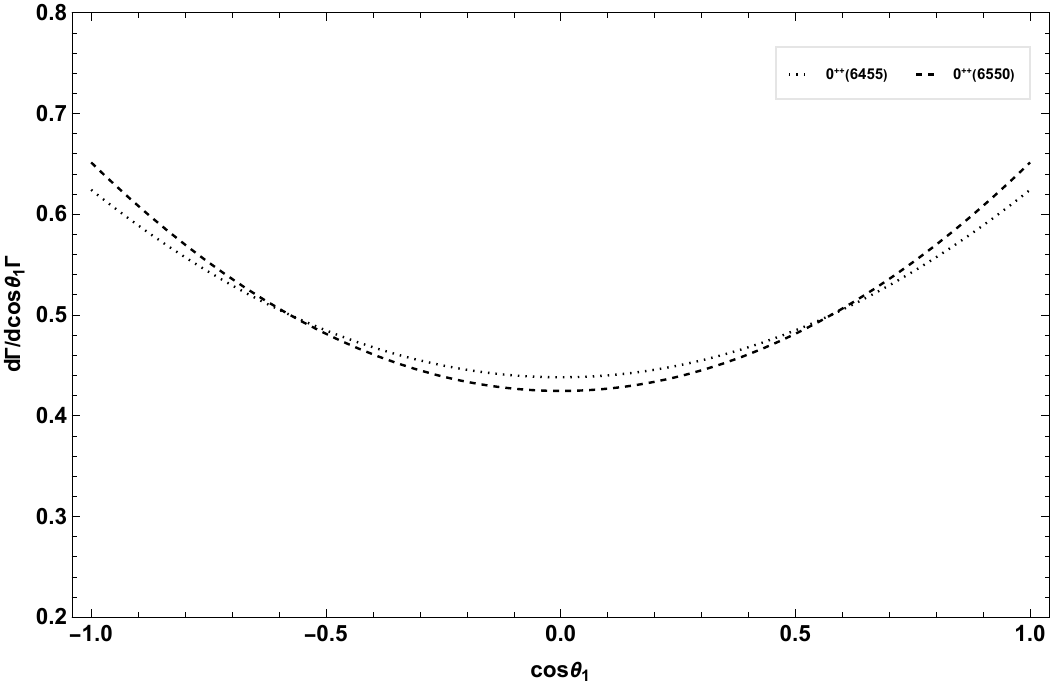}
	\includegraphics[width=0.45\textwidth]{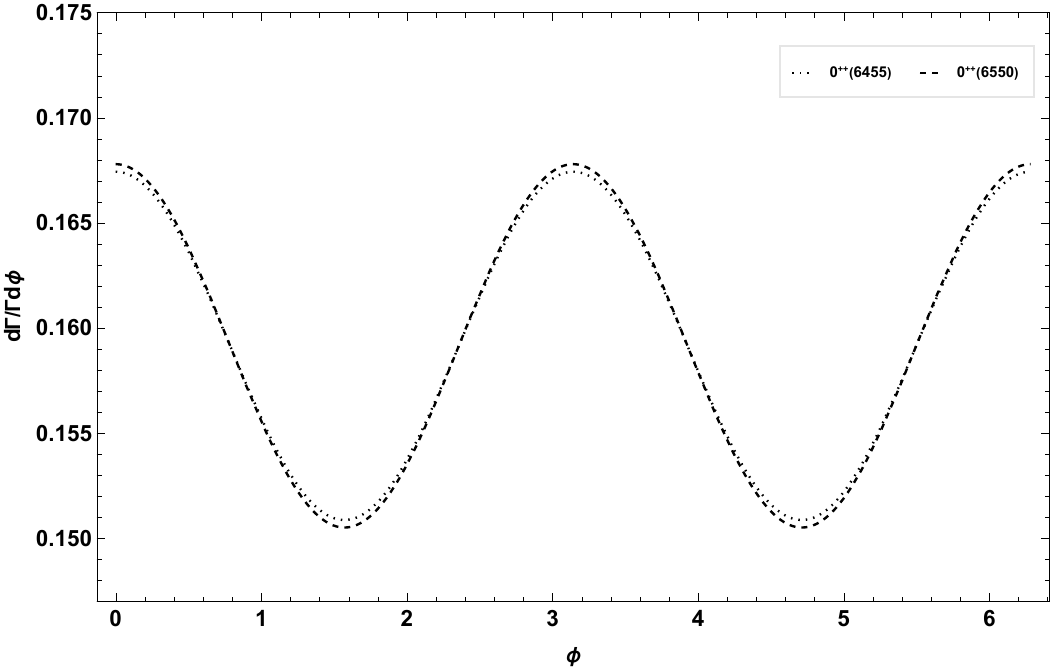}
	\caption{The $\theta_1$ and $\Phi$ distributions for various tetraquarks near 6.6GeV into double $J/\psi(\to \mu^++\mu^-)$ through quark models. }\label{Figthetadis1f}
\end{figure}
\begin{figure}[htp]
	\includegraphics[width=0.45\textwidth]{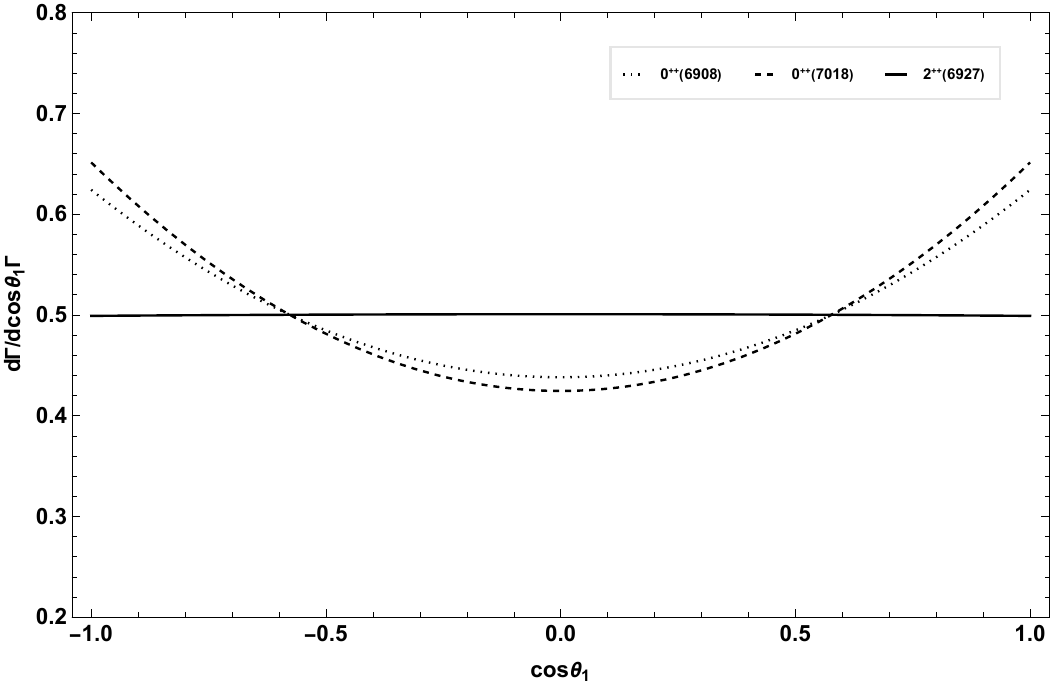}
	\includegraphics[width=0.45\textwidth]{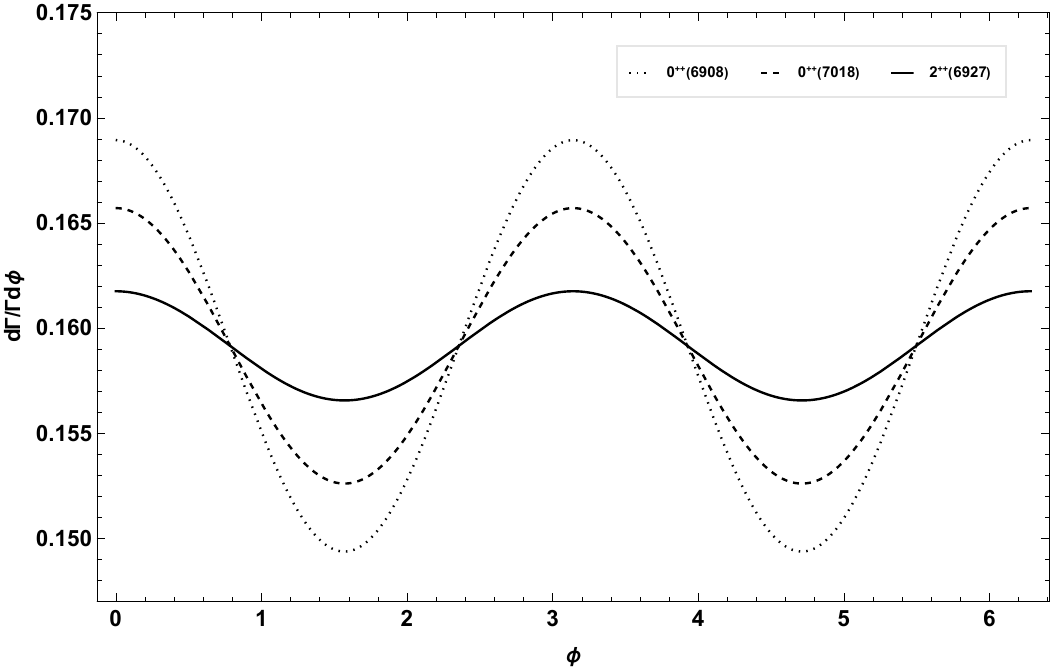}
	\caption{The $\theta_1$ and $\Phi$ distributions for various tetraquarks near 6.9GeV into double $J/\psi(\to \mu^++\mu^-)$ through quark models.}\label{Figthetadis2f}
\end{figure}
\begin{figure}[htp]
	\includegraphics[width=0.45\textwidth]{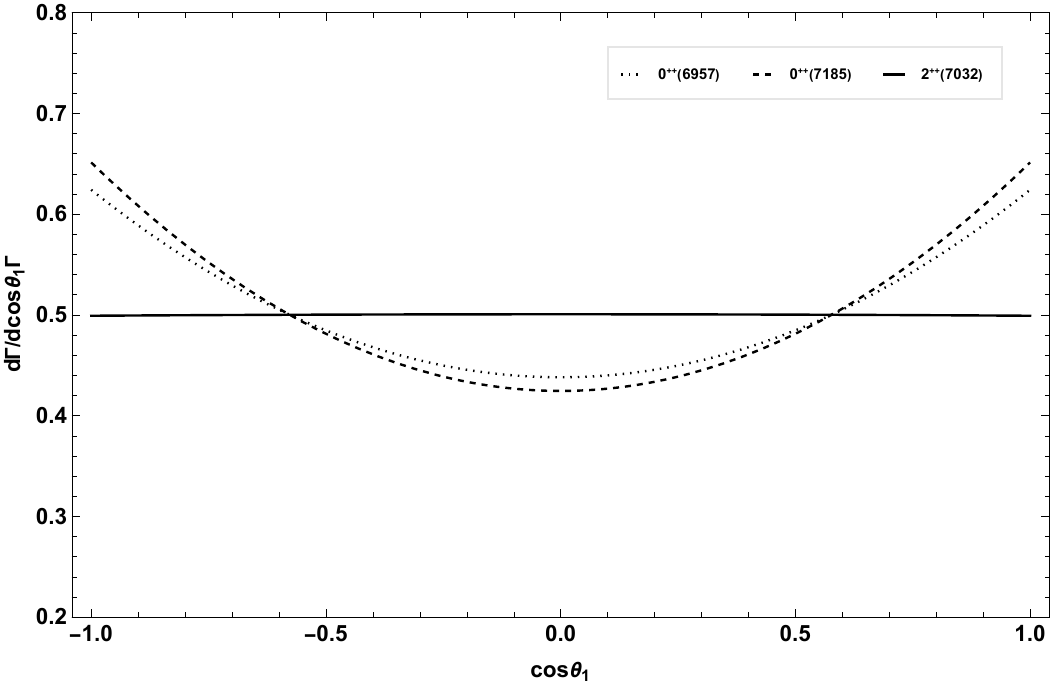}
	\includegraphics[width=0.45\textwidth]{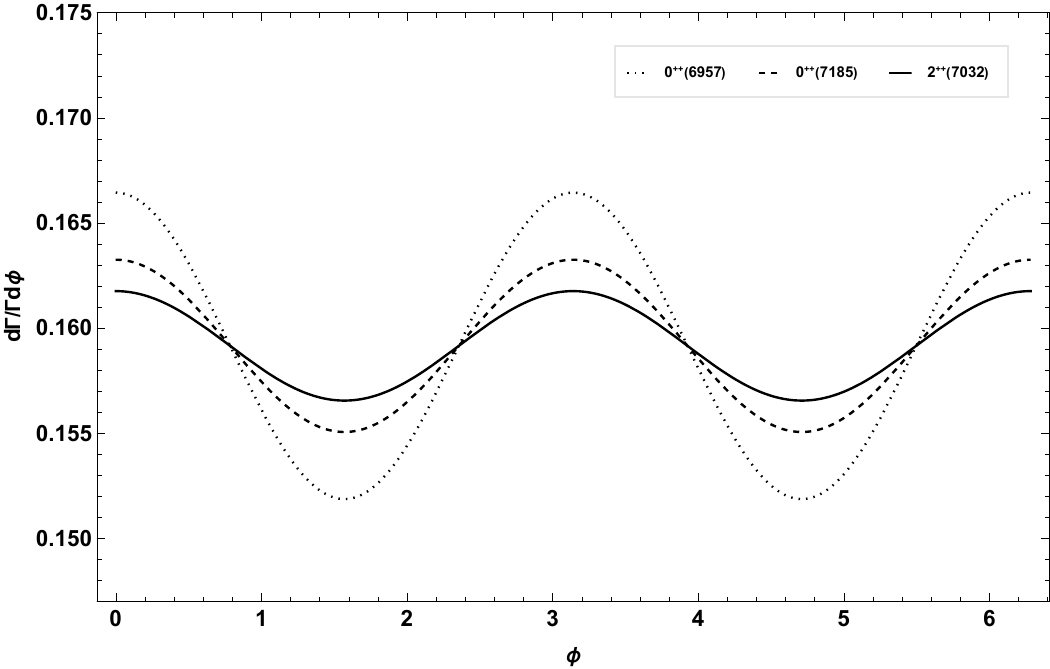}
	\caption{The $\theta_1$ and $\Phi$ distributions for various tetraquarks near 7.1GeV into double $J/\psi(\to \mu^++\mu^-)$ through quark models. }\label{Figthetadis3f}
\end{figure}

\begin{figure}[htp]
	\includegraphics[width=0.45\textwidth]{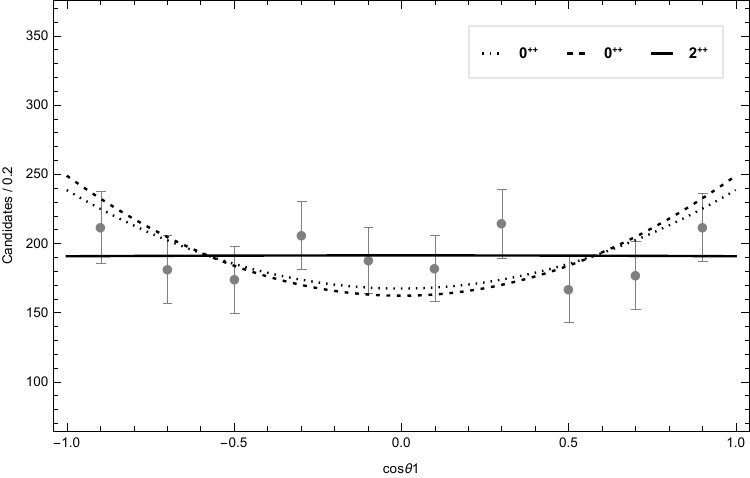}
	\includegraphics[width=0.45\textwidth]{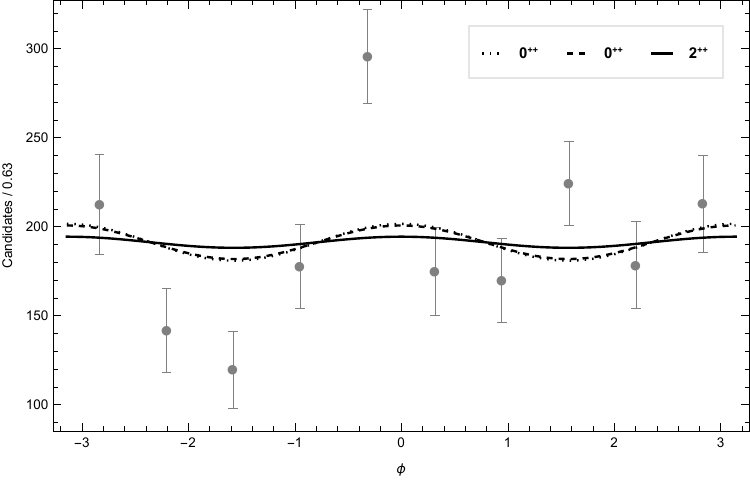}
	\caption{The Comparison  of $\theta_1$ and $\Phi$ distributions for the fully charm tetraquarks into double $J/\psi(\to \mu^++\mu^-)$ between the experiment and the theory. The experimental data with error bars are taken from the very recent CMS measurement~\cite{CMS:2025fpt}.}\label{compare}
\end{figure}

Taking into account the Pauli principle and color confinement for the four-quark system $cc\bar{c}\bar{c}$, we
have 4 configurations for $1S$-wave ground states,
and 12 configurations for the $2S$-wave radial excitations. The spin-parity quantum numbers, notations, and wave functions
for these configurations are presented in Table~\ref{ConventionI}. In the first column of the table, we used the notation of $n^{2S+1}L_{J}$. It should be noted that the table presents the wave functions corresponding to the mixed masses. These wave functions need to be linearly superposed to obtain the physical states. 
The wave function of the final state is obtained within the $J$-$J$
coupling scheme
\begin{eqnarray}
	\left|BC\right\rangle  & = & \chi_{11}^{SM_S} \varphi_{000}^{1}\varphi_{000}^{2}\phi\left|11\right\rangle . \label{eq:WFBC}
\end{eqnarray}
We use a plane wave to describe the relative motion
of the two mesons in the final state
\[
\phi=\frac{1}{\left(2\pi\right)^{3/2}}e^{-i\boldsymbol{p}_{f}\cdot(\boldsymbol{r}_{f_{1}}-\boldsymbol{r}_{f_{2}})},
\]
where $\boldsymbol{p}_{f}$ is the three-momentum of the mesons in
the final state, and $\boldsymbol{r}_{f_{1}}$ and $\boldsymbol{r}_{f_{2}}$
are the position coordinates of the hadrons 1 and 2 in the final state.
The assumption of a plane wave simplifies the treatment of the relative
motion between the mesons by treating them as free particles.

Calculating the matrix elements in color and spin space is relatively
straightforward. The integration of the spatial part is shown below.
For example, one of the spatial part of the integral is given by
\begin{eqnarray}
	&&\left\langle \varphi_{000}^{1}\left(\omega_{f_{1}}\right)\varphi_{000}^{2}\left(\omega_{f_{2}}\right)\phi\right|\hat{O}_{ij}\left|\psi_{000}^{1S}\left(\omega_{i}\right)\right\rangle \nonumber\\&=&I_{Nor}\int\hat{O}e^{-\sum_{i,j}A_{ij}\boldsymbol{\xi}_{i}\cdot\boldsymbol{\xi}_{j}}e^{-i\boldsymbol{p}_{f}\cdot\boldsymbol{\xi}_{3}}
	\left(Y_{00}\right)^{5}d^{3}\boldsymbol{\xi}_{1}d^{3}\boldsymbol{\xi}_{2}d^{3}\boldsymbol{\xi}_{3},\nonumber\\
\end{eqnarray}
where $\hat{O}_{ij}$ stands for the spatial-dependent operator, and $I_{Nor}$ is a normalization factor independent of the integration
variables. In
the calculations, the plane wave should be expanded as
\begin{equation}
	e^{i\boldsymbol{P}\cdot\boldsymbol{r}}=\sum_{l=0}^{\infty}\sqrt{4\pi(2l+1)}i^{l}j_{l}(Pr)Y_{l0}(\hat{\boldsymbol{r}}),
\end{equation}
where the momentum $P$ is assumed to be along the $z$ direction.
With the above steps, the integration of the spatial part can be obtained. Then, by applying the transformation relationship between the partial-wave and helicity amplitudes, we can obtain the final angular distributions.

Then, we present the numerical results for the angular distribution for double charmonia transition through quark model. We plot the  polar angle $\theta_1$ distribution and $\Phi$ distribution for various tetraquarks with different spin-parities near the 6.6 GeV, 6.9 GeV, 7.1GeV respectively. The values for their masses and spin-parity quantum numbers are from Ref.~\cite{liu:2020eha}. Note that the curve of the polar angle $\theta_2$ distribution is identical to the $\theta_1$ distribution due to the symmetry of final two vector mesons. These results are shown in Fig.~\ref{Figthetadis1f}-\ref{Figthetadis3f}.

Recently, the CMS Collaboratioin has released experimental measurements on the cascade decay of the fully charm tetraquark to $4\mu $~\cite{CMS:2025fpt}. Therefore, in order to compare the experimental data, we plotted the angular distribution results obtained by summing the contributions of the fully charm tetraquark states in three mass regions together with the experimental data points. The polar angle $\theta_1$ distribution for various tetraquarks with different spin-parities is shown in Fig.~\ref{compare}. The CMS data on the polar angle $\theta_1$ distribution are  well described by the $J^{PC}=2^{++}$ assignment from the quark model. The distribution of plane angle $\Phi$ between the two decay planes is also shown in Fig.~\ref{compare}. The figure reveals a $\cos(2\Phi)$ modulation in the $\Phi$ distribution, but the theoretical variation is smaller in amplitude than what is observed experimentally. Addressing this issue likely demands improved experimental precision in particle detection efficiency. In addition, the precise theoretial calculation is also required. We will study them in future works.

\subsection{Heavy Quark Spin Symmetry}
In this subsection, we introduce how to calculate the helicity amplitudes using heavy quark spin symmetry. In two-body decays of $T_{4c}\to D_{(s)}^*+\bar{D}_{(s)}^*$, there are two typical scales: the heavy quark mass $M$ and the nonperturbative scale $\Lambda_{QCD}$ in QCD. The short-distance and long-distance interactions can be factorized under the heavy quark limit with  $M\gg \Lambda_{QCD}$~\cite{Isgur:1989vq,Falk:1990yz}.  

The S-matrix for $T_{4c}\to H_{c\bar{q}}H^{(\prime)}_{q\bar{c}}$ at the order ${\cal O}(g_s^2)$ of strong coupling can be written as
\begin{align}\label{A1}
	&\langle H H'\vert S \vert T_{4c}\rangle \nonumber \\
	&=(-ig_s)^2\int\int  \mbox{d}^4 x  \mbox{d}^4 y \nonumber \\
	&\times\langle H H'|\mbox{T}A^{\mu}(x)A^{\nu}(y) j_\mu^\prime(x)j_\nu(y)|  T_{4c}\rangle+{\cal O}(g_s^4) \nonumber \\
	&=(-ig_s)^2C_AC_F\int\int  \mbox{d}^4 x  \mbox{d}^4 y \int\frac{ \mbox{d}^4 k}{(2\pi)^4}\frac{ ie^{-ik\cdot (x-y)}}{k^2+i0} \nonumber \\
	&\times\langle H H'|\mbox{T}j_\mu^\prime(x)j^\mu(y)|  T_{4c}\rangle+{\cal O}(g_s^4) \nonumber \\
	&=-g_s^2C_AC_F\int\int  \mbox{d}^4 x  \mbox{d}^4 y \int\frac{ \mbox{d}^4 k}{(2\pi)^4}\frac{ ie^{-ik\cdot (x-y)}}{k^2+i0} \nonumber \\
	&\times\langle H H'|\mbox{T}j_\mu^\prime(x)| n \rangle \langle n | j^\mu(y)|  T_{4c}\rangle+{\cal O}(g_s^4) \nonumber \\
	&=-g_s^2C_AC_F(2\pi)^4\delta^4(k+p_n-p_H-p_H^\prime)\nonumber \\
	&\times \delta^4(p_T-p_n-k)\int d^4k\frac{i}{k^2+i0} \nonumber \\
	&\times \langle H H'|\mbox{T}j_\mu^\prime(0)| n \rangle \langle n | j^\mu(0)|  T_{4c}\rangle+{\cal O}(g_s^4) \nonumber \\
	&=i(2\pi)^4\delta^4(P_T-p_H-p_H^\prime) \frac{-g_s^2C_AC_F}{(p_T-p_n)^2+i0} \nonumber \\
	&\times \langle H H'|\mbox{T}j_\mu^\prime(0)j^\mu(0)|  T_{4c}\rangle+{\cal O}(g_s^4) \nonumber \\
	&=i(2\pi)^4\delta^4(P_T-p_H-p_H^\prime)\mathcal{A}(T_{4c}\to H_{c\bar{q}}H^{(\prime)}_{q\bar{c}})
\end{align}
where the last line removing a factor $i(2\pi)^4\delta^4(P_T-P_H-P_{H'})$ will give the decay amplitude. $j_\mu(0)=\bar{c}\gamma_\mu c$ and $j_\mu^\prime(0)=\bar{q}\gamma_\mu q$ are the heavy vector current and light vector current ,respectively. $p_n$ is the momentum of an intermediate state. This intermediate state can be understood as the state after the four quarks that constitute the initial $T_{4c}$ emit a virtual gluon.

Then we introduce how to calculate the Feynman amplitude for process $T_{4c} \to D_s^{(*)} \bar{D}_s^{(*)}$. For the fully charmed tetraquark state, there are three forms of local operators in total. And these local operator can be expressed as~\cite{Liu:2025mxv},
\begin{align}
	\mathcal{O}_{\mathbf{6} \otimes \overline{\mathbf{6}}}^{(0)}  =&\left[\psi_a^T\left(i \sigma^2\right) \psi_b\right]\left[\chi_c^{\dagger}\left(i \sigma^2\right) \chi_d^*\right] \mathcal{C}_{\mathbf{6} \otimes \overline{\mathbf{6}}}^{a b ; c d},\nonumber \\
	\mathcal{O}_{\overline{\mathbf{3}} \otimes \mathbf{3}}^{(0)}  =&-\frac{1}{\sqrt{3}}\left[\psi_a^T\left(i \sigma^2\right) \sigma^i \psi_b\right]\left[\chi_c^{\dagger} \sigma^i\left(i \sigma^2\right) \chi_d^*\right] \mathcal{C}_{\overline{\mathbf{3}} \otimes \mathbf{3}}^{a b ; c d},\nonumber \\
	\mathcal{O}_{\overline{\mathbf{3}} \otimes \mathbf{3}}^{(2)}  =&\left[\psi_a^T\left(i \sigma^2\right) \sigma^m \psi_b\right]\left[\chi_c^{\dagger} \sigma^n\left(i \sigma^2\right) \chi_d^*\right]\mathcal{C}_{\overline{\mathbf{3}} \otimes \mathbf{3}}^{a b ; c d}\epsilon_{T}^{mn*}, \label{2}
\end{align}
where $\sigma^i$ is the Pauli matrix, $\chi$ and $\psi$ are two-dimensional spinor fields, and $\epsilon_T^{ij*}$ is the polarization tensor of the tetraquark state. The explicit expression for the color projection operator is
\begin{align}
	\mathcal{C}_{\mathbf{6} \otimes \overline{\mathbf{6}}}^{ab;cd} &= \frac{1}{2\sqrt{6}}\left(\delta_{ac}\delta_{bd}+\delta_{ad}\delta_{bc}\right),\nonumber\\
	\mathcal{C}_{\overline{\mathbf{3}} \otimes \mathbf{3}}^{ab;cd} &= \frac{1}{2\sqrt{3}}\left(\delta_{ac}\delta_{bd}-\delta_{ad}\delta_{bc}\right). 
\end{align} 

As shown by the local operators in Eq.~\ref{2}, when projecting the tetraquark state, we first project the two quarks and the two antiquarks into (anti)diquark pairs, and then further combine them into a fully charm tetraquark state with the corresponding quantum numbers. According to the Feynman diagram (b) in Fig.~\ref{Figmechanism}, the total amplitude can be written as follows,
\begin{align}
	i\mathcal{A}=&\frac{ig_s^2\xi_{D_s^{(*)}}}{P_g^2}\bar{u}(p_5)u(p_1)\bar{v}(p_2)\gamma^\mu u(p_3) \nonumber \\ 
	&\times \bar{v}(p_4)v(p_8)\bar{u}(p_7)\gamma_\mu v(p_6). 
\end{align}
In the above equation, $P_g$ is the momentum of the gluon propagator, The four quark momenta from top to bottom on the left side of the Feynman diagram are denoted as $p_1$, $p_2$, $p_3$, $p_4$ while the four quark momenta from top to bottom on the right side of the Feynman diagram are denoted as $p_5$, $p_6$, $p_7$, $p_8$. In order to project the initial and final states onto hadronic states with the corresponding quantum numbers, we can project the fermion chain into the following form,
\begin{align}
	i\mathcal{A}_1&=\int_{0}^1 dx \int_{0}^1 dx^\prime\frac{ig_s^2\xi_{D_s^{(*)}}}{P_g^2}\mathcal{C}^{ab;cd}_{6 \bigotimes \bar{6}}  \nonumber \\
	&\times Tr
	\Big[
	\Phi_{D_s^{(*)}}(x)\Pi_0\gamma^\rho\Pi_0\Phi_{\bar{D}_s^{(*)}}(x^\prime)\gamma_\rho
	\Big],   \nonumber \\
	i\mathcal{A}_2&=\int_{0}^1 dx \int_{0}^1 dx^\prime\frac{ig_s^2\xi_{D_s^{(*)}}}{P_g^2}J_0^{\mu\nu}\mathcal{C}^{ab;cd}_{\bar{3} \bigotimes 3} \nonumber \\
	&\times Tr
	\Big[
	\Phi_{D_s^{(*)}}(x)\Pi_{1,\mu}\gamma^\rho\Pi_{1,\nu}\Phi_{\bar{D}_s^{(*)}}(x^\prime)\gamma_\rho
	\Big], \nonumber \\
	i\mathcal{A}_3&=\int_{0}^1 dx \int_{0}^1 dx^\prime\frac{ig_s^2\xi_{D_s^{(*)}}}{P_g^2}\epsilon^{*\mu\nu}_{T_{4c}}\mathcal{C}^{ab;cd}_{\bar{3} \bigotimes 3} \nonumber \\
	& \times Tr
	\Big[
	\Phi_{D_s^{(*)}}(x)\Pi_{1,\mu}\gamma^\rho\Pi_{1,\nu}\Phi_{\bar{D}_s^{(*)}}(x^\prime)\gamma_\rho
	\Big],	\label{1}
\end{align}
where $\mathcal{A}_i$ corresponds to the three different projection operators of the tetraquark state. Because we consider the Feynman diagram (b) in Fig.~\ref{Figmechanism}, a factor $\xi_{D_s^{(*)}}$ needs to be introduced to balance the dimensions. This unknown coefficient cancels out upon normalization and does not affect our calculation of the angular distribution. In the Eq.~\ref{1}, the the spin-singlet and the spin-triplet projector for (anti)charm quark pair are
\begin{align}
	&\Pi_{0}=\frac{(\frac{\slashed{P}_T}{4}+m_c)\gamma_5}{\sqrt{2}}, \nonumber \\
	&\Pi_{1,\mu}=\frac{(\frac{\slashed{P}_T}{4}+m_c)\gamma_\mu}{\sqrt{2}}.
\end{align}
And the covariant projector $J_0^{\mu\nu}$ is
\begin{align}
	J_0^{\mu\nu}=\frac{1}{\sqrt{3}}(-g^{\mu\nu}+\frac{P_T^\mu P_T^\nu}{m_T^2}).
\end{align}
Meanwhile, the polarization tensor $\epsilon^{*\mu\nu}$ is defined as
where the polarization tensor is defined as,
\begin{align}
	&\epsilon_{\mu\nu}(\pm 2)=\epsilon_\mu(\pm) \epsilon_\nu(\pm), \nonumber \\
	&\epsilon_{\mu\nu}(\pm 1)=\frac{1}{\sqrt{2}}
	[\epsilon_\mu(\pm) \epsilon_\nu(0)+\epsilon_\mu(0) \epsilon_\nu(\pm)], \nonumber \\
	&\epsilon_{\mu\nu}(0)=\frac{1}{\sqrt{6}}
	[\epsilon_\mu(+) \epsilon_\nu(-)+\epsilon_\mu(-) \epsilon_\nu(+)]+\sqrt{\frac{2}{3}}\epsilon_\mu(0) \epsilon_\nu(0).
\end{align}

\begin{figure}[htp]
	\includegraphics[width=0.45\textwidth]{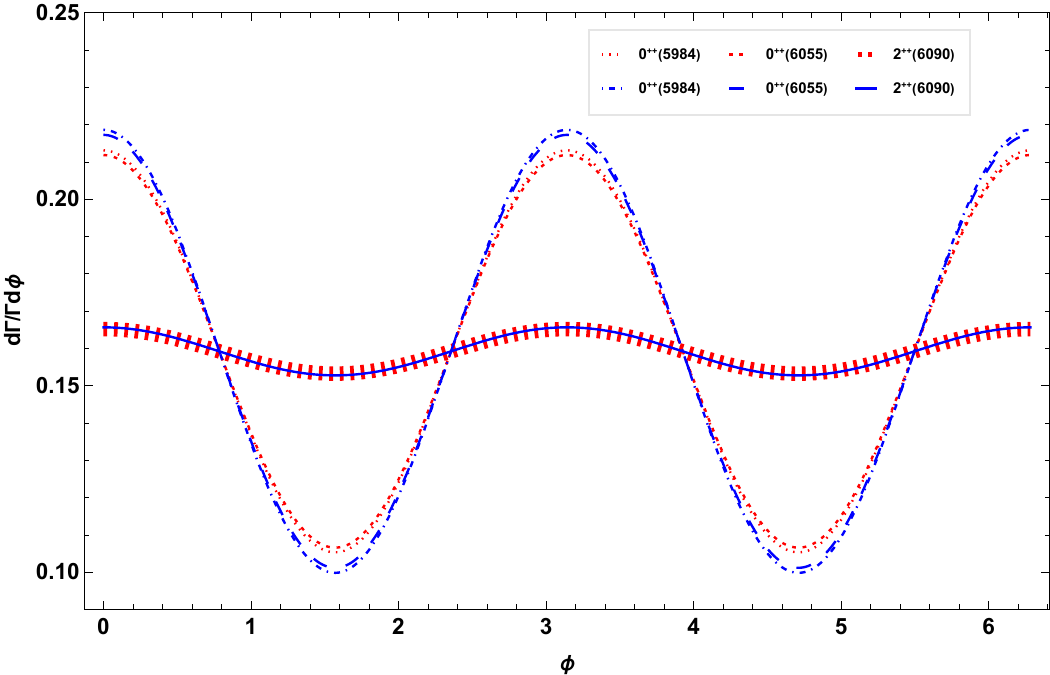}
	\includegraphics[width=0.45\textwidth]{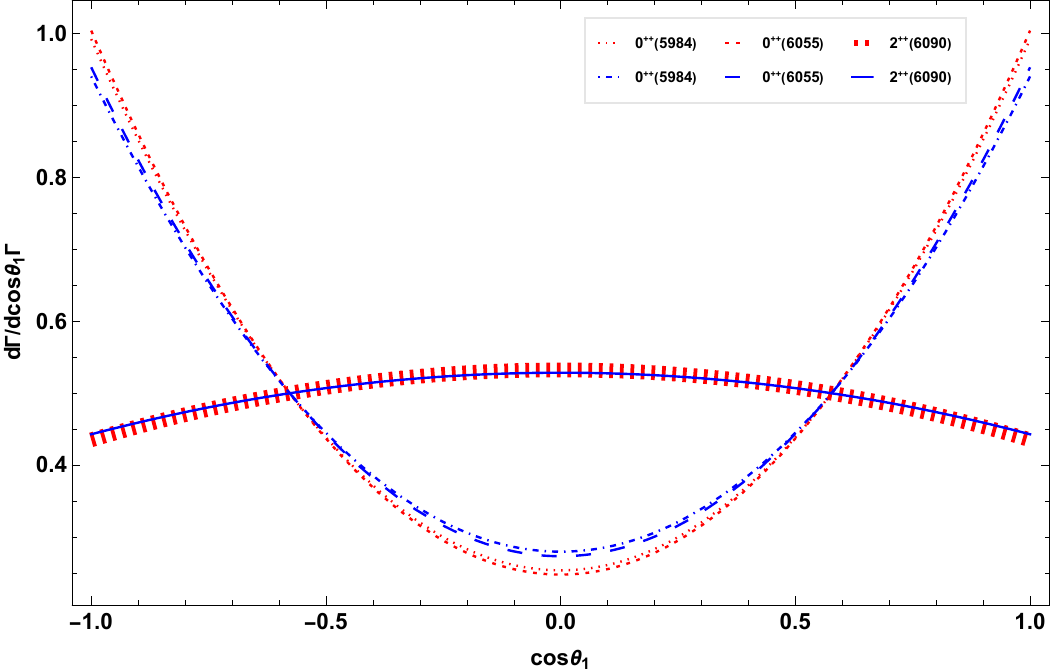}
	\caption{The $\theta_1$ and $\Phi$ distributions for various tetraquarks near 6.1 GeV into $D_{(s)}^*(\rightarrow D_{(s)} \pi)$ and $\bar{D}_{(s)}^*(\rightarrow \bar{D}_{(s)} \pi)$ using heavy quark spin symmetry. The red curves indicate the decay channels involving $D^*$ and blue curves indicate the decay channels involving $D_s^*$.}\label{FigthetaphiDDsstar}
\end{figure}

\begin{figure}[htp]
	\includegraphics[width=0.45\textwidth]{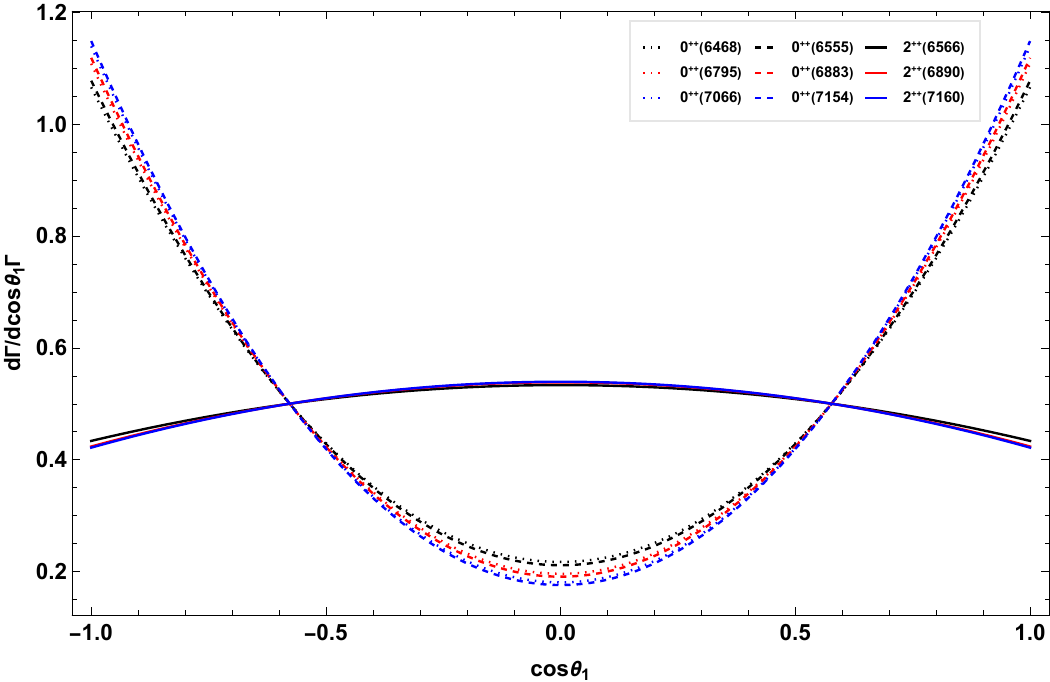}
	\includegraphics[width=0.45\textwidth]{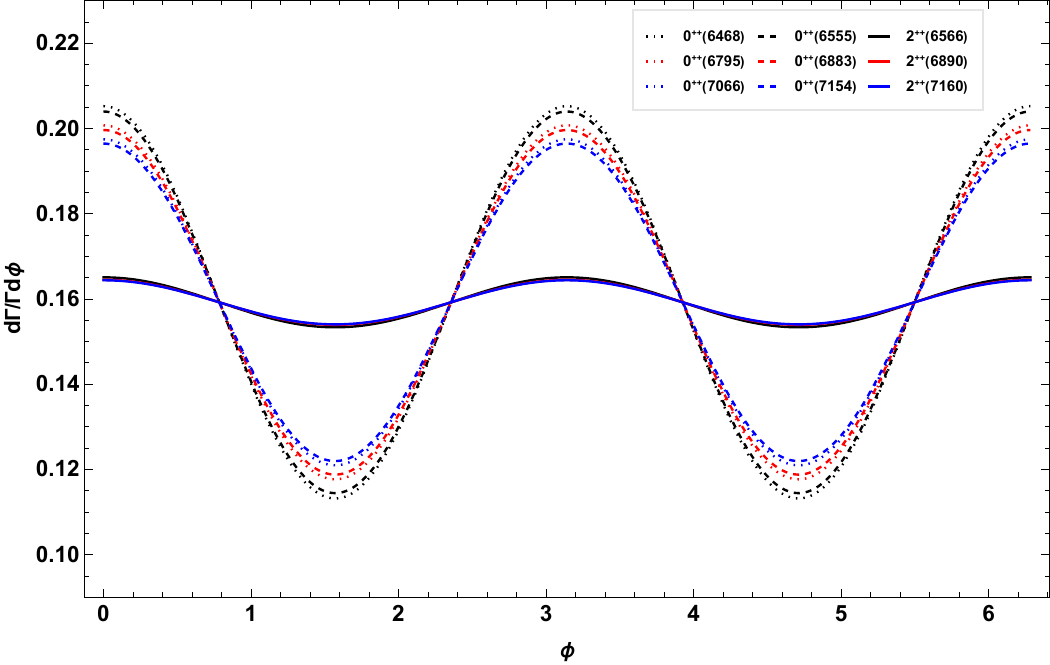}
	\caption{The $\theta_1$ and $\Phi$ distributions for various tetraquarks around 6.6 GeV, 6.9 GeV and 7.1 GeV into $D^*(\rightarrow D \pi)$ and $\bar{D}^*(\rightarrow \bar{D} \pi)$ using heavy quark spin symmetry. }\label{FigthetadisDstar}
\end{figure}

\begin{figure}[htp]
	\includegraphics[width=0.45\textwidth]{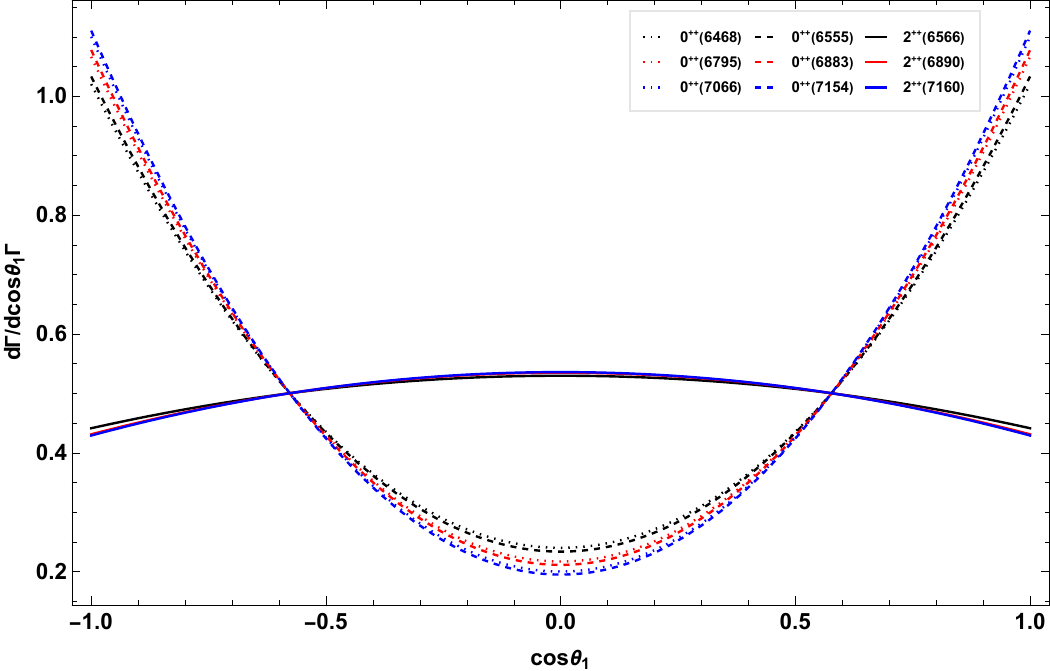}
	\includegraphics[width=0.45\textwidth]{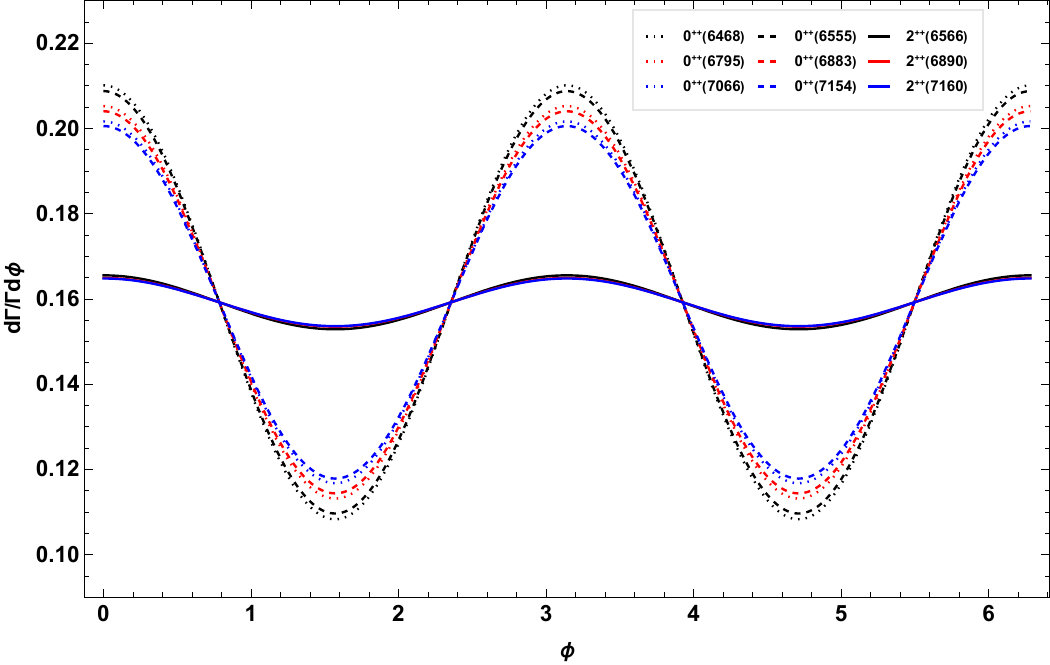}
	\caption{The $\theta_1$ and $\Phi$ distributions for various tetraquarks around 6.6 GeV, 6.9 GeV and 7.1 GeV into $D_s^*(\rightarrow D_s \pi)$ and $\bar{D}_s^*(\rightarrow \bar{D}_s \pi)$ using heavy quark spin symmetry. }\label{FigthetadisDSstar}
\end{figure}

$\phi_{D_s^{(*)}}$ is the LCDA of the $D_s^{(*)}$ meson, which can be expressed as
\begin{align}
	\Phi_{D_s}(P_{D_s},x)&=\frac{i}{\sqrt{2N_c}}\gamma^5(\slashed{P}_{D_s}+m_{D_s})\phi_{D_s}(x), \nonumber \\
	\Phi_{D_s^*}(P_{D_s^*},x)&=-\frac{i}{\sqrt{2N_c}}
	\Big[
	\slashed{\epsilon}_L(\slashed{P}_{D_s^*}+m_{D_s^*})\phi_{D_s^*}^L(x) \nonumber \\
	&+\slashed{\epsilon}_T(\slashed{P}_{D_s^*}+m_{D_s^*})\phi_{D_s^*}^T(x)
	\Big],
\end{align}
where $\slashed{\epsilon}_{L,T}$ denotes the longitudinal and transverse polarization vectors of the $D^*_s$ meson. Because in the case we consider there is no momentum exchange between the upper and lower quarks, the momentum fraction carried by the final-state light quark is completely determined. Therefore, the amplitudes of the two tetraquark states with different spin quantum numbers can ultimately be simplified to
\begin{align}
	i\mathcal{A}^0& \propto \mathcal{C}^{ab;cd}_{6 \bigotimes \bar{6}} Tr
	\Big[
	\Pi^{D_s^{(*)}}\Pi_0\gamma^\rho\Pi_0\Pi^{\bar{D}_s^{(*)}}\gamma_\rho
	\Big]   \nonumber \\
	&+J_0^{\mu\nu}\mathcal{C}^{ab;cd}_{\bar{3} \bigotimes 3} Tr
	\Big[
	\Pi^{D_s^{(*)}}\Pi_{1,\mu}\gamma^\rho\Pi_{1,\nu}\Pi^{\bar{D}_s^{(*)}}\gamma_\rho
	\Big],  \nonumber \\
	i\mathcal{A}^2& \propto \epsilon^{\mu\nu*}_T\mathcal{C}^{ab;cd}_{\bar{3} \bigotimes 3} Tr
	\Big[
	\Pi^{D_s^{(*)}}\Pi_{1,\mu}\gamma^\rho\Pi_{1,\nu}\Pi^{\bar{D}_s^{(*)}}\gamma_\rho
	\Big], \label{5}
\end{align}
where the project operator of $D_s^{(*)}$ can be expressed as
\begin{align}
	&\Pi^{D_s}=\gamma^5(\slashed{P}_{D_s}+m_{D_s}), \nonumber \\
	&\Pi^{D_s^*}=\slashed{\epsilon}^*_{D_s^*}(\slashed{P}_{D_s}+m_{D_s}).
\end{align}
By simplifying Eq.~\ref{5}, we can extract the relative magnitudes of the coefficients given in Eq.~\ref{3} and Eq.~\ref{4}, thereby obtaining the normalized helicity amplitudes.

Then we give the numerical result of  fully charm tetraquark into $D^*\bar{D}^*$ and $D_s^*\bar{D}_s^*$. In the heavy quark spin symmetry analysis, the values for the masses and spin-parity quantum numbers of fully charm tetraquarks are from Ref.~\cite{Zhu:2020xni}. Since the angular distribution shapes of these two decay channels in different mass ranges are very similar, we will analyze them using $T_{4c} \to D_{(s)}^*\bar{D}_{(s)}^*$ near 6.1 GeV as an example. The results are plotted in Fig.~\ref{FigthetaphiDDsstar}. We find that the dependence of plane angle $\Phi$  for double charm mesons channel are similar to that for double $J/\psi$ channel, while the dependence of polar angle  $\theta_1$ is completely different due to the difference of final products. Especially we find the curves of the spin-two state remain nearly constant with respect to the angle $\Phi$, exhibiting an independence characteristic. Meanwhile, we find that for the spin-0 tetraquark state, the variation of the 
$\theta_1$ angular distribution is significantly larger than that for the spin-2 tetraquark state. These phenomena may provide a supplementary plan for determining the spin-parity of double $J/\psi$ exotic states.
To achieve these goals, one may consider the process, $T_{4c}\to {D^*}(2010)^+(\to D^{0}+\pi^+)+{D^*}(2010)^-(\to \bar{D}^{0}+\pi^-)$ along with $D^{0}\to K^-+\pi^+$, to be a favorable observation channel. Additionally, we believe this channel can be used to probe fully charm tetraquark below the double charmonia threshold that have not yet been observed in experiment. The remaining angular distribution numerical results in different mass region are presented in Fig.~\ref{FigthetadisDstar}-\ref{FigthetadisDSstar}.

\subsection{Model-independent extraction from experiment}

Through the CMS measurements of the angular distributions of existing processes, we can also extract information about the helicity amplitudes. Since the two kinds of decay angular distributions currently available are not yet sufficient to extract all the helicity amplitudes, we redefine the parameters on which the angular distributions depend, in the hope of obtaining reliable estimates of these parameters.

We can rewrite the angular distribution formula to fit the experimental data and determine the parameters therein. For the angular distribution of $T_{4c} \to J/\psi J/\psi\to 4\mu$, we can rewrite it as 
\begin{align}
	&\frac{d\Gamma}{\Gamma d\cos\theta_1}=\frac{1}{2}(1+\xi \frac{3\cos^2\theta_1-1}{2}), \nonumber \\
	&\frac{d\Gamma}{\Gamma d\Phi}=\frac{1}{2\pi}(1+\beta \cos2\Phi).
\end{align}
For the spin-0 tetraquark state, the two new parameters in terms of helicity amplitudes are,
\begin{align}
	&\xi_0=\frac{|F_{11}|^2-|F_{00}|^2}{\mathcal{N}_0},  \quad \beta_0=\frac{|F_{11}|^2}{8\mathcal{N}_0},
\end{align}
	with
	\begin{align}
	&\mathcal{N}_0=2|F_{11}|^2+|F_{00}|^2.
\end{align}
And for the spin-2 tetraquark state, the two new parameters in terms of helicity amplitudes are,
\begin{align}
	&\xi_2=\frac{-|F_{00}|^2+|F_{11}|^2-|F_{10}|^2+|F_{11}|^2}{\mathcal{N}_2},  \quad \beta_2=\frac{|F_{11}|^2}{2\mathcal{N}_2},
\end{align}
	with
	\begin{align}
	&\mathcal{N}_2=|F_{00}|^2+2|F_{11}|^2+2|F_{1-1}|^2+4|F_{10}|^2.
\end{align}

\begin{figure}[htbp]
	\centering
	\begin{subfigure}{0.45\textwidth}
		\centering
		\includegraphics[width=1.05\linewidth]{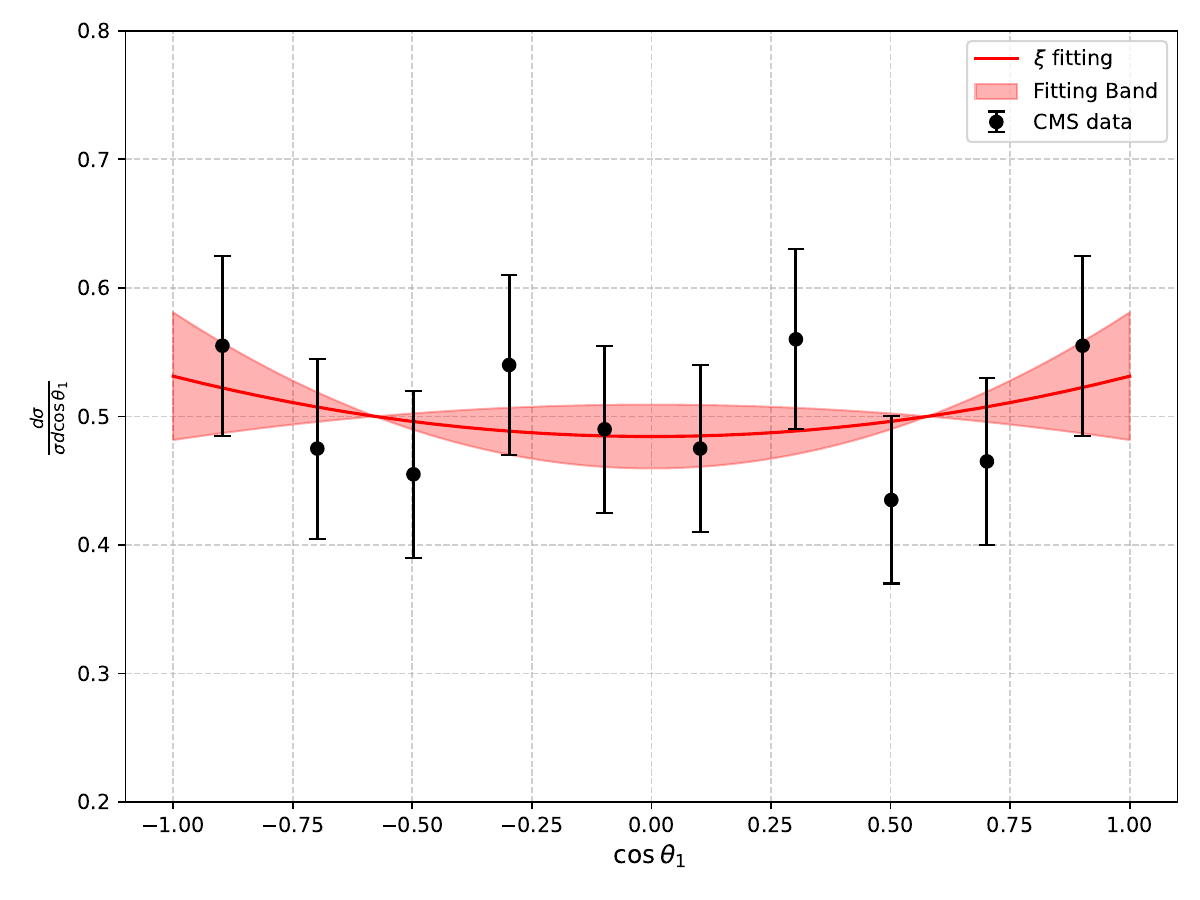}
	\end{subfigure}
	\hfill
	\begin{subfigure}{0.45\textwidth}
		\centering
		\includegraphics[width=1.05\linewidth]{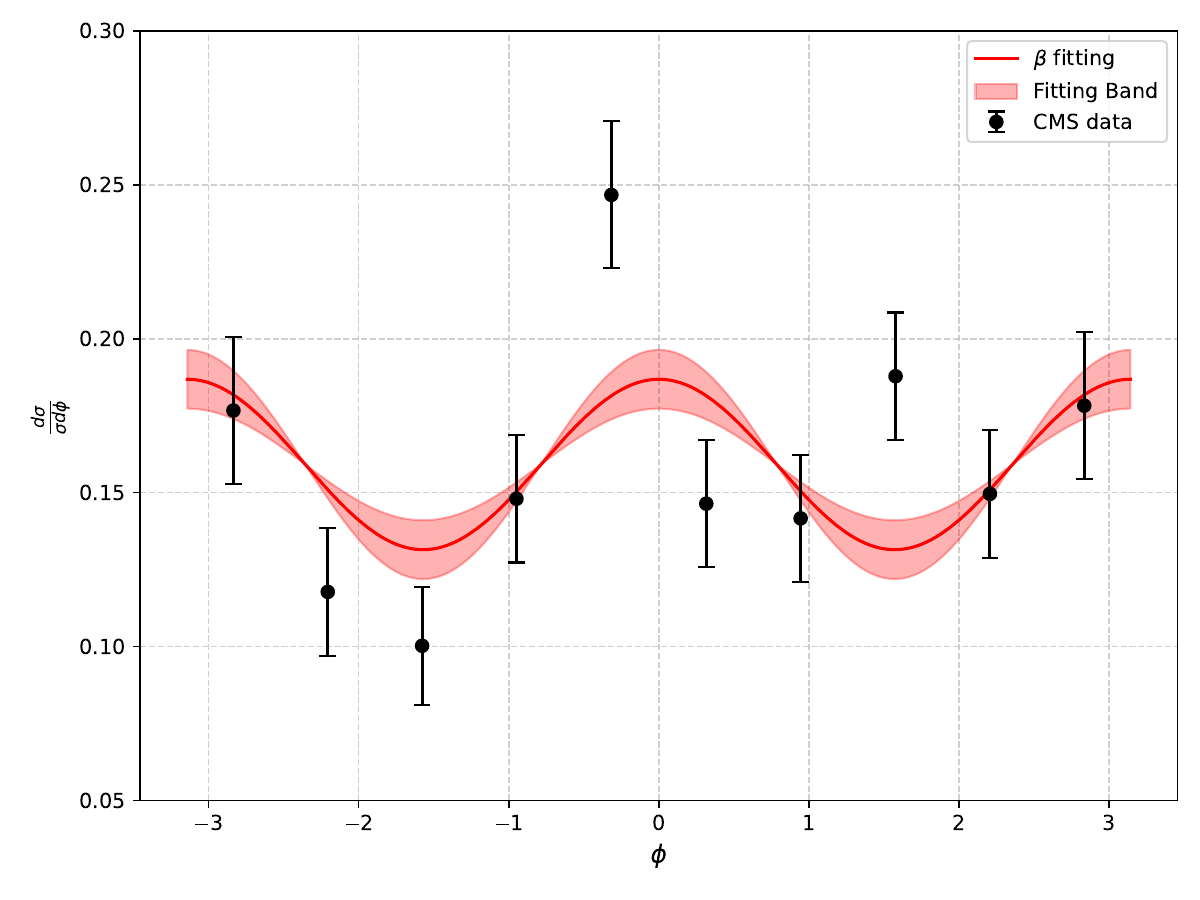}
	\end{subfigure}
	\caption{The angular distribution fitting for $\xi$ and $\beta$. The experimental data with error bars are taken from the very recent CMS measurement~\cite{CMS:2025fpt}.}
	\label{fig2}
\end{figure}

It can be seen that the fitting formulae for these two cases are identical. The optimal fitted parameter values obtained based on the experimental data are $\xi=0.063 \pm 0.099$ with the $\chi^2/dof=0.44$ and $\beta=0.174 \pm 0.060$ with the $\chi^2/dof=2.6$. The fitting results are shown in Fig.~\ref{fig2}. The two distributions show different levels of agreement with the fitted expressions. The small $\chi^{2}/\mathrm{dof}$ for the $\cos\theta_1$ distribution indicates that the data are consistent with the fitted curve, but their uncertainties leave $\xi$ weakly constrained. In contrast, the $\phi$ distribution gives a relatively large $\chi^{2}/\mathrm{dof}$. As seen in Fig.~\ref{fig2}, several bins at approximately symmetric values of $\phi$ have substantially different yields, whereas the $\cos 2\phi$ term predicts equal yields at those angles. These deviations dominate the fit residuals and reduce the quality of the determination of $\beta$. 

\section{Quantum Entanglement Analysis}

One characteristic property of fully charm tetraquark decay into double vector mesons is the quantum correlation between the final two vector mesons. We assume that, without considering the initial state polarization, the final two vector meson system is a pure state. Then, in the center-of-mass frame of $T_{4c}$, the spin wave function of the vector meson pair can be expressed as
\begin{align}
	| \Psi \rangle= \frac{1}{\sqrt{N^2}}\sum_{\lambda_1\lambda_2}F^J_{\lambda_1\lambda_2} | V(\lambda_1)V(\lambda_2) \rangle,
\end{align}
where $N$ is the normalization factor which is equal to $\sum_{ij}h_{ij}^{ij}$. The density matrix of the vector meson pair can be expressed as $\rho=| \Psi \rangle \langle \Psi |$.

\begin{figure*}[htp]
	\centering
	\includegraphics[width=0.32\textwidth]{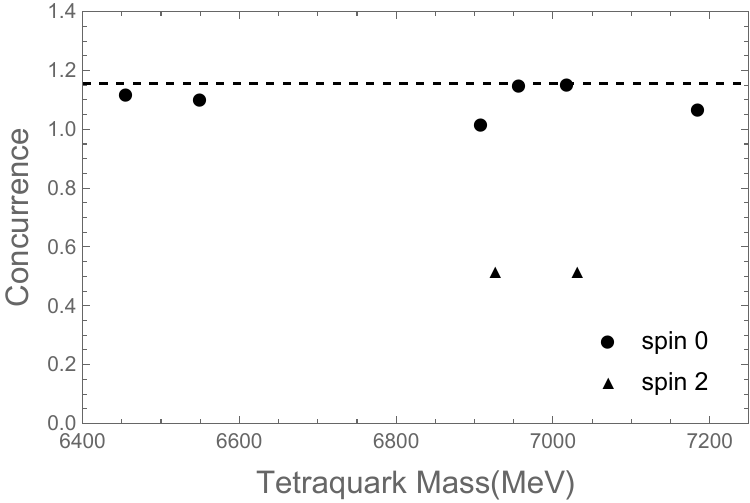}%
	\hfill
	\includegraphics[width=0.32\textwidth]{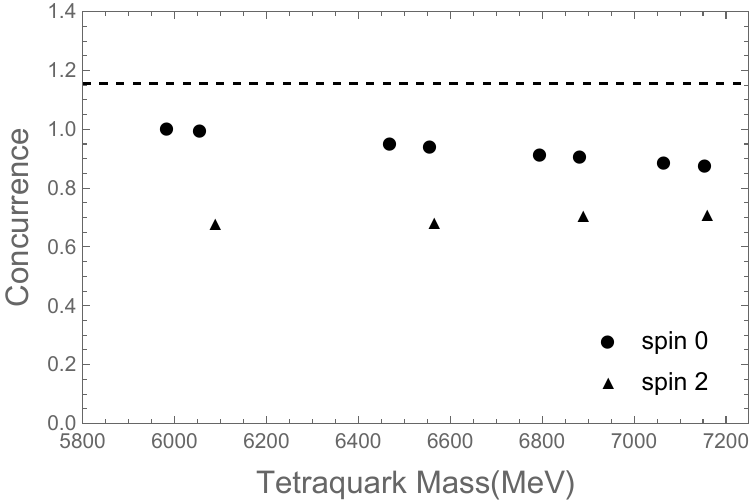}%
	\hfill
	\includegraphics[width=0.32\textwidth]{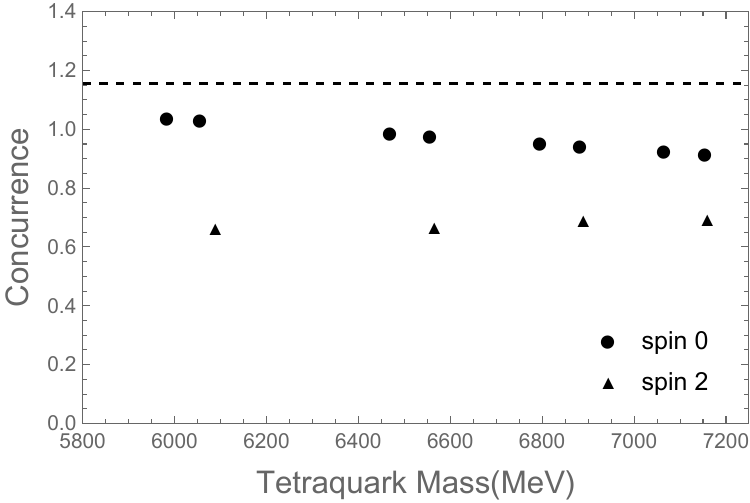}%
	\caption{From left to right, they respectively represent the consistent numerical results of concurrence for $J/\psi J/\psi$, $D^*\bar{D}^*$, and $D_s^*\bar{D}_s^*$ obtained from the corresponding models. }\label{FigConcurrence}
\end{figure*}

Typically, to describe the quantum entanglement effects in a two-particle system, various entanglement measures can be introduced, such as the concurrence and  von Neumann entropy~\cite{Chen:2024syv}. Here, we use concurrence to characterize the degree of quantum entanglement in fully charm tetraquark  decay process. For a pure state of two qutrits, the concurrence can be computed as~\cite{Barr:2024djo}
\begin{align}
	\mathcal{C}=\sqrt{2(1-\mathrm{Tr}\rho_A^2)},
\end{align}
where $\rho_A$ represents the partial trace of the total density matrix over one of the subsystems (denoted as A). For the two qutrit system, its value satisfies $[max\left(0,LB\right),\frac{2}{\sqrt{3}}]$~\cite{Morales:2023gow,Zhao2010,Eltschka2015}. The lower bound of concurrence can be expressed as
\begin{align}
	LB=\sqrt{\frac{1}{3}}\Big[(\sum_i \lambda_i)^2-1\Big],
\end{align}
where $\lambda_i$ is the square root of the eigenvalues of $\rho_A$. The larger the value, the stronger the degree of entanglement. To solve the quantum correlation, it is necessary to calculate the reduced density matrix first. For the tetraquark with an initial spin of 0, the reduced density matrix is expressed as
\begin{align}
	\rho_A=\frac{1}{N}
	\begin{pmatrix}
		|F_{-1-1}|^2 & 0 & 0 \\
		0 & |F_{00}|^2 & 0 \\
		0 & 0 & |F_{11}|^2
	\end{pmatrix}.	
\end{align}
And for the tetraquark with spin 2, $\rho_A$ matrix can be expressed as
\begin{align}
	\rho_A=\frac{1}{N}
	\begin{pmatrix}
		A_1 & A_2 & A_3 \\
	    A_4 & A_5 & A_6 \\
		A_7 & A_8 & A_9\\
	\end{pmatrix},
\label{rhoA}
\end{align}	
where 
\begin{align}
	& A_1=|F_{-1-1}|^2+|F_{-10}|^2+|F_{-11}|^2, \nonumber\\
	& A_2=F_{-1-1}F_{0-1}^*+F_{-10}F_{00}^*+F_{-11}F_{01}^*, \nonumber\\
	& A_3=F_{-1-1}F_{1-1}^*+F_{-10}F_{10}^*+F_{-11}F_{11}^*, \nonumber\\
	& A_4=F_{0-1}F_{-1-1}^*+F_{00}F_{-10}^*+F_{01}F_{-11}^*, \nonumber\\
	& A_5=|F_{0-1}|^2+|F_{00}|^2+|F_{01}|^2, \nonumber\\
	& A_6=F_{0-1}F_{1-1}^*+F_{00}F_{10}^*+F_{01}F_{11}^*, \nonumber\\
	& A_7=F_{1-1}F_{-1-1}^*+F_{10}F_{-10}^*+F_{11}F_{-11}^*, \nonumber\\
	& A_8=F_{1-1}F_{0-1}^*+F_{10}F_{00}^*+F_{11}F_{01}^*, \nonumber\\
	& A_9=|F_{1-1}|^2+|F_{10}|^2+|F_{11}|^2.
\end{align}	
By substituting the specific numerical value of the helicity amplitude, the concurrence of the two vector mesons in the final state can be obtained. We plot the concurrence of fully charm tetraquark into double charmonia transition and double charmed meson transition in Fig.~\ref{FigConcurrence}. Whether for $T_{4c} \to J/\psi J/\psi $ channel or $T_{4c} \to D_{(s)}^*  \bar{D}_{(s)}^*$ channel, the entanglement of the meson pair obtained from the decay of the spin-2 tetraquark state is lower than that in the case of decay from the spin-0 tetraquark state. These results will provide a different perspective on understanding the tetraquark structures and their decay behaviors.

Conversely, if we consider quantum entanglement as a fundamental principle in the decays, we can derive the following conclusions. The upper and lower limits on concurrence will constrain the helicity amplitudes.  The upper and lower limits on concurrence will constrain the helicity amplitudes. For the tetraquark with spin 0, we have the constraint on helicity amplitudes
\begin{align}
	&max\left(0, \sqrt{\frac{1}{3}} \left [\frac{(\sqrt{h_{00}^{00}}+2\sqrt{h_{11}^{11}})^2}{N}-1\right]\right) \nonumber \\
	&\leq \sqrt{2(1-\frac{\left(h_{00}^{00}\right)^2+2\left(h_{11}^{11}\right)^2}{N^2})} \leq \frac{2}{\sqrt{3}}.
\end{align}
For the tetraquark with spin 2, we have the constraint
\begin{align}
	&max\left(0, LB_2\right)\leq (2-2\left(\frac{h_{00}^{00}+2h_{01}^{01}}{N}\right)^2 \nonumber \\
	&-4\left(\frac{h_{10}^{10}+h_{11}^{11}+h_{1-1}^{1-1}}{N}\right)^2-8\left(\frac{h^{10}_{00}+h^{11}_{01}+h^{1-1}_{01}}{N}\right)^2 \nonumber \\
	& -4\left(\frac{h^{10}_{10}+h^{1-1}_{11}+h^{11}_{1-1}}{N}\right)^2)^\frac{1}{2} \leq \frac{2}{\sqrt{3}}. 
\end{align} 
The above result has been simplified using Eqs.~\ref{relation1} and \ref{relation2}. $LB_2$ is determined by the three eigenvalues of Eq.~\ref{rhoA}. These three eigenvalues can be expressed as
\begin{widetext}
\begin{align}
	&\lambda_1=\frac{1}{3}B_0+\frac{\sqrt[3]{B_1+\sqrt{B_2}}}{{3 \sqrt[3]{2}}}-\frac{\sqrt[3]{2} \left(-A_1^2+A_5 A_1+A_9 A_1-A_5^2-A_9^2-3 A_2A_4-3 A_3 A_7-3 A_6 A_8+A_5 A_9\right)}{3 \sqrt[3]{B_1+\sqrt{B_2}}}, \\
	&\lambda_2=\frac{1}{3}B_0-\frac{\left(1-i \sqrt{3}\right) \sqrt[3]{B_1+\sqrt{B_2}}}{6 \sqrt[3]{2}}\nonumber\\&~~~~~~~~+\frac{\left(1+i \sqrt{3}\right) \left(-A_1^2+A_5 A_1+A_9 A_1-A_5^2-A_9^2-3 A_2 A_4-3 A_3 A_7-3 A_6 A_8+A_5 A_9\right)}{32^{2/3} \sqrt[3]{B_1+\sqrt{B_2}}}, \\
	&\lambda_3=\frac{1}{3}B_0-\frac{\left(1+i \sqrt{3}\right) \sqrt[3]{B_1+\sqrt{B_2}}}{6\sqrt[3]{2}}\nonumber\\&~~~~~~~~+\frac{\left(1-i \sqrt{3}\right) \left(-A_1^2+A_5 A_1+A_9 A_1-A_5^2-A_9^2-3 A_2 A_4-3 A_3 A_7-3 A_6 A_8+A_5 A_9\right)}{3\ 2^{2/3} \sqrt[3]{B_1+\sqrt{B_2}}}.
\end{align}	
\end{widetext}
In the above formula, the expression for $B_i$ are
\begin{widetext}
\begin{align}
	&B_0=A_1+A_5+A_9, \\
	&B_1=2A_1^3-3 A_5 A_1^2-3 A_9 A_1^2-3 A_5^2 A_1-3 A_9^2 A_1+9 A_2 A_4 A_1+9 A_3 A_7 A_1-18 A_6 A_8 A_1\nonumber \\&~~~~~~~~+12A_5 A_9 A_1+2 A_5^3+2 A_9^3-3 A_5 A_9^2 +9 A_2 A_4 A_5-18 A_3 A_5 A_7+27 A_2 A_6 A_7+27 A_3 A_4 A_8\nonumber \\
	&~~~~~~~~+9 A_5 A_6 A_8-3 A_5^2 A_9-18 A_2 A_4 A_9+9 A_3 A_7 A_9+9 A_6A_8 A_9,\\
	&B_2=4(-A_1^2+A_5 A_1+A_9 A_1-A_5^2-A_9^2-3 A_2 A_4-3 A_3 A_7-3 A_6 A_8+A_5 A_9){}^3+(2 A_1^3-3 A_5 A_1^2-3 A_9 A_1^2-3 A_5^2 A_1 \nonumber \\
	&~~~~~~-3 A_9^2A_1+9 A_2 A_4 A_1+9 A_3 A_7 A_1-18 A_6 A_8 A_1+12 A_5 A_9 A_1+2 A_5^3+2 A_9^3-3 A_5 A_9^2+9 A_2 A_4 A_5-18 A_3 A_5 A_7 \nonumber \\
	&~~~~~~+27 A_2 A_6 A_7+27 A_3 A_4 A_8+9 A_5 A_6A_8-3 A_5^2 A_9-18 A_2 A_4 A_9+9 A_3 A_7 A_9+9 A_6 A_8 A_9){}^2.
\end{align}	
\end{widetext}
We also provided the most general constraints in the case of symmetry breakdown, so that they can be applied to more general situations. The constraint brought by entanglement in the most general case for the tetraquark with spin 0 is
\begin{widetext}
	\begin{align}
		max\left(0, \sqrt{\frac{1}{3}} \left [\frac{(\sqrt{h^{-1-1}_{-1-1}}+\sqrt{h^{11}_{11}}+\sqrt{h^{00}_{00}})^2}{N}-1\right]\right)\leq \sqrt{2(1-\frac{\left(h_{00}^{00}\right)^2+\left(h_{11}^{11}\right)^2+\left(h_{-1-1}^{-1-1}\right)^2}{N^2})} \leq \frac{2}{\sqrt{3}}.
	\end{align}
\end{widetext}
And the constraint of the entanglement range on the tetraquark with spin 2 can be expressed as
\begin{widetext}
\begin{align}
	&max\left(0, LB_2\right)\leq (2-2\left(\frac{h_{00}^{00}+h_{01}^{01}+h_{0-1}^{0-1}}{N}\right)^2-2\left(\frac{h_{10}^{10}+h_{11}^{11}+h_{1-1}^{1-1}}{N}\right)^2
	-2\left(\frac{h_{-10}^{-10}+h_{-11}^{-11}+h_{-1-1}^{-1-1}}{N}\right)^2 \nonumber \\
	&-4\left(\frac{|h^{10}_{00}+h^{11}_{01}+h^{1-1}_{0-1}|}{N}\right)^2-4\left(\frac{|h^{-10}_{00}+h^{-11}_{01}+h^{-1-1}_{0-1}|}{N}\right)^2-4\left(\frac{|h^{10}_{10}+h^{-11}_{11}+h^{-1-1}_{1-1}|}{N}\right)^2)^\frac{1}{2} \leq \frac{2}{\sqrt{3}}.
\end{align}	
\end{widetext}
All these constraints may assist experimentalists in fitting the angular distribution curves using experimental data.

\section{Conclusion}

The confirmation of the completely new family of fully charm tetraquark states is expected to play a significant role in testing the Standard Model of particle physics and advancing our understanding of the color confinement mechanism in QCD. In this work, we have  for the firt time  studied the helicity amplitudes for the fully charm tetraquark states cascade decays  $T_{4c}\to J/\psi(D_{(s)}^*)+J/\psi(\bar{D}_{(s)}^*)$ along with $J/\psi\to \mu^++\mu^-$ or $D_{(s)}^*\to D_{(s)}+\pi$, showing that spin-0 and spin-2 states can be distinguished.  
The quantum entanglement effect between the two vector mesons from fully charm tetraquark is also studied, showing that the concurrence is slightly changed for the decays of various radially excited tetraquark with identical $J^{PC}$, and the  difference is reduced for higher excited tetraquark with different spin.  We also find that the spin-0 state decay to  vector-meson pair exhibits a higher degree of quantum entanglement than that in spin-2 state decays.  The constraints formula for the helicity amplitudes are given assuming the fundamental principle of entanglement, which shall be useful for experimentalists to fit the helicity amplitudes. These results of angular distribution can be tested in current particle physics experiments such as LHCb, ATLAS, CMS and Belle-II.

It is worth noting that the decays into open charm meson pair such as $D^{(*)}_{(s)}\bar{D}_{(s)}^{(*)}$  are crucial for searching for fully charm tetraquark states below the double charmonia threshold, as these processes are the main decay modes in this case.  The processes mentioned above will undoubtedly provide a fresh perspective on double $J/\psi$ exotic states, helping to reveal the multi-peak structures, clarify their internal properties, and establish fully charm tetraquark family.

\begin{acknowledgments}
{\it Acknowledgement.}
We thank Prof. F.-K. Guo for useful discussion. This work is supported by the National Natural Science Foundation of China (Grants No.~12322503, No.~12675120, No.~12047503, No.12235018, No.12175065, No.E411645Z10,  No.1A2024000016), and National Key Basic Research Program of China under Contract No. 2020YFA0406300,
Q. Zhao and B.-S. Zou are also supported in part, by the DFG and NSFC funds to the Sino-German CRC 110 ``Symmetries and the Emergence of
Structure in QCD'' (NSFC Grant No. 12070131001, DFG Project-ID 196253076), and Strategic Priority Research Program of Chinese
Academy of Sciences (Grant No. XDB34030302).
\end{acknowledgments}

\end{document}